%% file: main.tex
\documentclass[11pt,letterpaper]{article}
\pdfoutput=1

\usepackage{amsmath,amssymb,amsfonts,mathrsfs,amsthm}
\usepackage[]{graphicx}
\usepackage{dsfont}

\usepackage{xcolor}
\definecolor{darkblue}{rgb}{0.1,0.1,.7}
\usepackage[colorlinks, linkcolor=darkblue, citecolor=darkblue, urlcolor=darkblue, linktocpage]{hyperref} 
\usepackage[square, comma, compress,numbers]{natbib}
\usepackage{geometry}
\usepackage[margin=10pt,font=small,labelfont=bf]{caption}

\numberwithin{equation}{section}

\def\mbb{\mathbb}
\def\mbf{\mathbf}
\def\mca{\mathcal}

\def\mrm{\mathrm}
\def\msc{\mathscr}

\def\msf{\mathsf}

\def\th{\tfrac{1}{2}}
\def\half{\frac{1}{2}}
\def\pd{\partial}
\def\Oo{\mathcal{O}}
\def\sO{\mathrm{O}}
\def\l{\ell} 
\def\ba{\boldsymbol{a}}
\def\bb{\boldsymbol{b}}
\def\qaq{\quad \text{and} \quad}
\def\qor{\quad \text{or} \quad}

\def\lra{\leftrightarrow}
\def\a{\alpha}
\def\b{\beta}

\def\dd{\delta}

\def\la{\lambda}

\def\ga{\gamma}

\def\DD{\Delta}

\def\vareps{\varepsilon}
\def\phib{\bar{\phi}}

\newcommand{\expec}[1]{\langle #1 \rangle}

\newcommand{\reef}[1]{(\ref{#1})}
\def\beq{\begin{equation}} 
\def\eeq{\end{equation}}
\def\nn{\nonumber} 
\def\bsub{\begin{subequations}}
\def\esub{\end{subequations}}
\def\ldef{\mathrel{\mathop:}=}
\def\rdef{=\mathrel{\mathop:}}

\newcommand{\NO}[1]{{:\!#1\!:}}
\newcommand{\norm}[1]{\lvert #1 \rvert}
\newcommand{\ud}[2]{^{#1}_{\phantom{#1}#2}}
\newcommand{\du}[2]{_{#1}^{\phantom{#1}#2}}

\newcommand{\floor}[1]{\left \lfloor #1 \right \rfloor }

\theoremstyle{plain}

\newtheorem*{fact}{Fact}
\newtheorem*{thm}{Theorem}
\theoremstyle{remark}

\begin{document}

\vspace*{-.6in} \thispagestyle{empty}
\begin{flushright}
\end{flushright}
\vspace{1cm} {\Large
\begin{center}
  {\bf Bounds on multiscalar CFTs in the $\vareps$ expansion}
\end{center}}
\vspace{1cm}
\begin{center}
{\bf Matthijs Hogervorst${}^{a}$, Chiara Toldo${}^{b,c,d}$ }\\[2cm] 
{\it\small
  ${}^a$ Fields and Strings Laboratory, \'{E}cole Polytechnique F\'{e}d\'{e}rale de Lausanne (EPFL)\\
\vspace{2mm}
${}^b$ Centre de Physique Th\'eorique  (CPHT), Ecole Polytechnique, \\ 
91128 Palaiseau Cedex, France \\
\vspace{2mm}
${}^c$ Institut de Physique Th\'eorique, Universit\'e Paris Saclay, CEA, \\ CNRS, Orme des Merisiers, 91191 Gif-sur-Yvette Cedex, France \\ 
\vspace{2mm}
${}^d$ Institute for Theoretical Physics, University of Amsterdam, Science Park 904,\\  Postbus 94485, 1090 GL, Amsterdam, The Netherlands
}
\\
\end{center}
\vspace{14mm}

\begin{abstract}
  We study fixed points with $N$ scalar fields in $4 - \varepsilon$ dimensions to leading order in $\varepsilon$ using a bottom-up approach. We do so by analyzing $O(N)$ invariants of the quartic coupling $\la_{ijkl}$ that describes such CFTs. In particular, we show that $\la_{iijj}$ and $\la_{ijkl}^2$ are restricted to a specific domain, refining a result by Rychkov and Stergiou. We also study averages of one-loop anomalous dimensions of composite operators without gradients. In many cases, we are able to show that the $O(N)$ fixed point maximizes such averages. In the final part of this work, we generalize our results to theories with $N$ complex scalars and to bosonic QED. In particular we show that to leading order in $\varepsilon$, there are no bosonic QED fixed points with $N < 183$ flavors.
  
\end{abstract}
\vspace{12mm}

\newpage

{
\setlength{\parskip}{0.05in}
\tableofcontents
}


\newpage

\section{Introduction}
\label{sec:intro}
\input{sections/intro.tex}


\section{Review of the epsilon expansion}
\label{sec:review}
\input{sections/review.tex}


\section{Bounds on $\la_{ijkl}$ at fixed points}
\label{sec:boundsMAIN}
\input{sections/bounds.tex}


\section{Anomalous dimensions}
\label{anom-dim}
\input{sections/anomalous.tex}


\section{Bounds for complex scalars and bosonic QED}
\label{complex-scalar}
\input{sections/complex.tex}


\section{Discussion}
\label{sec:discussion}
\input{sections/discussion.tex}


\begin{appendix}
\input{sections/appendix.tex}

\end{appendix}


{\footnotesize
  
\bibliography{biblio}
\bibliographystyle{utphys.bst}

}


\end{document}

%% file: sections/intro.tex

Second-order phase transitions are described by scale-invariant (or rather conformally invariant) quantum field theories. There are two fundamentally different ways to approach the study of phase transitions. On the one hand, one can try to construct {specific} fixed points, using the logic of effective field theory. That is to say that one writes down a Hamiltonian for the order parameters of the system, including all possible relevant operators that are consistent with its global symmetry group. Given an effective Hamiltonian, one has to analyze whether it admits one or several fixed points that can describe the phase transition of interest. 

An orthogonal way of thinking of this problem is bottom-up: one could attempt to make a list of \emph{all} conformal field theories together with their spectrum (i.e.\@ their critical exponents). Given such a list, it would be possible to look for theories with a specific global symmetry group in two or three spacetime dimensions. The problem of classifying CFTs is known as the \emph{conformal bootstrap}~\cite{Ferrara:1973yt,Polyakov:1974gs,Rattazzi:2010gj} --- see Ref.~\cite{Poland:2018epd} for a comprehensive review.

Although both approaches appear to be radically different, they lead to the same experimental predictions. For instance, the Ising CFT in three dimensions has been studied for decades via renormalization group methods, yielding predictions for its critical exponents up to four significant digits~\cite{Pelissetto:2000ek}. In recent years, these predictions were recovered and even significantly improved by establishing rigorous bounds on the space of unitary, $\mbb{Z}_2$-symmetric CFTs with two relevant perturbations~\cite{El-Showk:2014dwa,Kos:2016ysd}.

Unfortunately, there is at present no simple method to classify all CFTs, even after fixing a spacetime dimension $d$ and a symmetry group $G$.  To make progress, we analyze in this paper a simplified classification problem that has been studied since the 1970s~\cite{Wilson:1971dc,Brezin:1973jt,Michel:1981yx,Michel:1983in,Michel:1985pq}. To be precise, we consider $N$ scalars $\phi^i$ that interact via a general quartic interaction,
\beq
\label{eq:Vphi}
V(\phi) = \la_{ijkl}  \, \phi^i \phi^j \phi^k \phi^l.
\eeq
without imposing any constraints on the couplings $\la_{ijkl}$, except for the fact that they must be real-valued (otherwise, the interaction~\reef{eq:Vphi} can flow to a non-unitary CFT).  In two or three dimensions, any fixed point obtained by turning on the interaction~\reef{eq:Vphi} is strongly coupled. However, in $d=4-\vareps$ dimensions with $\vareps \ll 1$, the RG flow of the couplings $\{\la_{ijkl}\}$ can be studied perturbatively. The beta functions of the $\la_{ijkl}$ are of the form
\beq
\label{eq:beta}
\beta(\la_{ijkl}) = - \vareps \la_{ijkl} + P_{ijkl}(\la)
\eeq
where $P_{ijkl}(\la)$ is an analytic function of the couplings that starts at order $\lambda^2$. Consequently, Eq.~\reef{eq:beta} has roots that can be expanded order by order in the small parameter $\vareps$. To study CFTs of this type, one can initially discard two- and higher-loop contributions to the beta function~\reef{eq:beta}.

At this stage, it is worth mentioning that the study of CFTs in the epsilon expansion is an active field of research. Let us briefly sketch some different approaches, without attempting to be exhaustive. Conformal symmetry strongly restricts perturbative fixed points, to the extent that certain anomalous dimensions can be determined using representation theory ideas~\cite{Rychkov:2015naa,Roumpedakis:2016qcg,Gliozzi:2017gzh}.  Moreover, the $\vareps$ expansion lends itself well to analytic bootstrap methods~\cite{Gopakumar:2016cpb,Alday:2017zzv,Henriksson:2018myn,Carmi:2020ekr}. For the Wilson-Fisher CFT, modern tools have been developed to compute anomalous dimensions systematically without Feynman diagrams~\cite{KEHREIN1993669,Liendo:2017wsn}. In other recent work~\cite{Codello:2018nbe}, equations of motion were combined with conformal symmetry to provide many quantitative results to leading order in $\vareps$. We also point to Ref.~\cite{Chai:2020zgq}, which recently discussed thermal phase transitions for a large class of multiscalar theories of the form~\reef{eq:Vphi}. 

The problem of classifying one-loop fixed points in the $\vareps$ expansion is mathematically well-defined,  but it is highly non-trivial: it consists of finding zeroes of $d_N$ quadratic polynomials in $d_N$ variables, where $d_N$ is the number of couplings $\la_{ijkl}$ (for instance, $d_2 = 5$ and $d_3 = 15$). Even for the case of $N=2$ fields, the classification of fixed points was completed only recently, by Osborn and Stergiou~\cite{Osborn:2017ucf}. For $N \geq 3$ however, no such classification is known. Progress has been made in a different direction: in Ref.~\cite{Rychkov:2018vya}, Rychkov and Stergiou were able to show that the norm $|\la|^2 = \la_{ijkl}^2$ of fixed point couplings is bounded from above. This development is interesting because it rules out large swaths of theory space $\mca{T}$, the $d_N$-dimensional space of couplings. Since $d_N \approx \tfrac{1}{24} N^4$, already for moderate values of $N$ it's extremely complicated to find fixed points inside the high-dimensional space $\mca{T}$. By showing that fixed points are instead restricted to a compact domain $\mca{D} \subset \mca{T}$, the task of hunting for interacting CFTs becomes much easier.\footnote{See for example the recent work~\cite{Codello:2020lta}, which describes an effort to look for fixed points numerically using the aforementioned bound on $\la_{ijkl}^2$.} The main goal of this work is to further shrink the allowed region $\mca{D}$ where CFTs can live.

This paper is organized as follows. In Sec.~\ref{sec:review}, we review the problem of finding one-loop fixed points with $N$ scalars in the epsilon expansion, and we discuss some families of known fixed points and their symmetry groups. For $N=3,4$ we provide a list of known CFTs along with quantitative data about their couplings $\la_{ijkl}$. In Sec.~\ref{sec:boundsMAIN}, we define $O(N)$ invariants of $\la_{ijkl}$, such as $\la_{iijj}$, and in turn we prove upper and lower bounds on these invariants. After that, in Sec.~\ref{anom-dim} we consider the spectrum of order-$\vareps$ anomalous dimensions $\ga^{(1)}$ of operators of the type $\Oo_r \sim \phi^{i_1}  \dotsm \phi^{i_r}$. A key result is that their averages $\expec{\ga}_r$ and $\expec{\ga^2}_r$ are strongly constrained --- in particular, we prove that the $O(N)$ fixed point maximizes these averages for many values of $N$ and $r$. In the final part of this manuscript, the above strategy is applied to theories with $N$ complex scalars as well as to bosonic QED. In particular, we show that there are no QED fixed points with fewer than 183 flavors. Several appendices provide proofs and explicit computations.

\textbf{Note:} During the completion of this paper we became aware of~\cite{Osborn:2020vya} which obtains a number of new fixed points for $N = 5,6,7$ via numerical and analytical methods, and which derives further bounds on $O(N)$ invariants that are consistent with ours.

%% file: sections/review.tex

In this section, we will briefly review the classification problem at hand. As announced in the introduction, we consider a theory of $N$ real scalars $\phi^i$ with a $\mbb{Z}_2$ symmetry $\phi^i \mapsto -\phi^i$ described by the Lagrangian
\beq
\label{eq:genLag}
\msc{L} = \half (\pd_\mu \phi^i)^2 + \frac{B}{4!}\, \la_{ijkl} \phi^i \phi^j \phi^k \phi^l + \text{counterterms}
\eeq
we have introduced an arbitrary factor $B > 0$ to normalize the couplings $\la_{ijkl}$. 
Without imposing any further restrictions, the coupling $\la_{ijkl}$ can be any symmetric tensor. In $d=4-\vareps$ dimensions with $\vareps$ small, the quartic couplings have mass dimension $\vareps \ll 1$ so they are weakly relevant. Working in minimal subtraction, the beta function for the couplings reads
\beq
\label{eq:MSbetaf}
\beta(\la)_{ijkl} = - \vareps \la_{ijkl} + \frac{B}{16\pi^2}\!\left( \la_{ijmn} \la_{klmn} + \la_{ikmn} \la_{jlmn} + \la_{ilmn} \la_{jkmn} \right) + \sO(\la^3).
\eeq
Cubic and higher-order terms in $\lambda$ can be computed in perturbation theory, by taking into account two- and higher-loop Feynman diagrams. In this scheme, the small parameter $\vareps$ only appears in the leading term $-\vareps \la_{ijkl}$: all higher-order terms are independent of $\vareps$. From now on, we will set $B=16\pi^2$ to simplify further formulas.

Consequently, we can look for fixed points $\beta(\la^\star) = 0$ of the form
\beq
\label{eq:laStar}
\la^\star_{ijkl}(\vareps) = \la^{}_{ijkl} \, \vareps +  \la^{(2)}_{ijkl} \, \vareps^2 + \ldots.
\eeq
The leading-order contribution $\la_{ijkl}$ is necessarily a root of the one-loop beta function
\beq
\label{eq:oneloopbeta}
\beta^{(1)}(\la)_{ijkl} = - \la_{ijkl} +  \la_{ijmn} \la_{klmn} + \la_{ikmn} \la_{jlmn} + \la_{ilmn} \la_{jkmn}.
\eeq
Terms $\la^{(n)}_{ijkl}$ that are subleading in $\vareps$ can be determined by taking into account $n$-loop contributions to the beta function. At least when $N=1$, these subleading contributions are completely fixed\footnote{A sufficient condition for this to happen for $N \geq 2$ is found in~\cite{Michel:1985pq}.} by the leading piece $\la_{ijkl}$. 

It is therefore an interesting problem to classify solutions of the one-loop beta function~\reef{eq:oneloopbeta}. Once such a one-loop CFT $\la_{ijkl}$ is found, it's possible to compute its spectrum of anomalous dimensions (again, to leading order in $\vareps$), which gives a rough estimate for its critical exponents in $d=3$, and its global symmetry group $G$ can be determined as well. If more precise quantitative predictions for 3$d$ physics are needed, it becomes necessary to perform a higher-loop computation and to resum its results.

\subsection{Global symmetries \label{glob-symm}}

By construction, any Lagrangian~\reef{eq:genLag} has a $\mbb{Z}_2$ global symmetry $\phi^i \mapsto -\phi^i$. A coupling $\la_{ijkl}$ can however have a larger global symmetry $G \subset O(N)$. In particular, there exists a maximally symmetric fixed point with an $O(N)$ global symmetry:
\beq
\label{eq:ONfp}
V(\phi) = \la_{ijkl} \, \phi^i \phi^j \phi^k \phi^l =  \frac{3}{N+8}  \, (\phi  \cdot \phi)^2.
\eeq
At the same time, there exist CFTs with discrete symmetry groups, like the \emph{hypercubic fixed point}:\footnote{For $N=4$, the potential~\reef{eq:cubAns} is nothing but the $O(4)$ model from~\reef{eq:ONfp}.}
\beq
\label{eq:cubAns}
V(\phi) = \frac{1}{N} \, (\phi \cdot \phi)^2 + \frac{N-4}{3N} \sum_{i=1}^N \phi_i^4.
\eeq
The above potential is invariant under $G = \mbb{Z}_2^N \rtimes \mca{S}_N$, where $\mca{S}_N$ permutes the $\phi^i$ and the $k$-th copy of $\mbb{Z}_2$ flips the sign of $\phi^k$. Finally, there are fixed points with symmetry groups that are products of discrete and continuous groups. Consider for instance
\beq
\label{eq:Ising}
\la_{ijkl} = \frac{1}{3} u_i u_j u_k u_l
\quad
\Leftrightarrow
\quad
V(\phi) = \frac{1}{3} (u \cdot \phi)^4
\eeq
where $u_i$ is an arbitrary unit vector. This is the well-known {Wilson-Fisher} CFT, which in two and three dimensions describes the Ising model. Its symmetry group is $\mbb{Z}_2 \times O(N-1)$, where $O(N-1)$ acts on fields orthogonal to $u^i$. Concretely, we can e.g.\@ set $u^i = (1,0,\ldots,0)$ such that the potential only depends on a single field $\phi^1$. It is easy to generalize to $k \leq N$ copies of the Wilson-Fisher fixed point:
 \beq
\label{eq:kIsing}
 V(\phi) = \frac{1}{3} \sum_{\a = 1}^k (u^{(\a)} \cdot \phi)^4.
 \eeq
where the $u^{(\a)}_i$ are unit vectors obeying $u^{(\a)} \cdot u^{(\b)} = \dd^{\a\b}$. The CFT~\reef{eq:kIsing} is invariant under $G = (\mbb{Z}_2^k \rtimes \mca{S}_k ) \times O(N-k)$.

Mathematically, the symmetry group of a coupling $\la_{ijkl}$ is its {stabilizer} under the action of $O(N)$. Concretely, an element $\mca{R}$ of the orthogonal group acts as
\beq
\la_{ijkl} \mapsto \la^{\mca{R}}_{ijkl} = \mca{R}\ud{i'}{i} \mca{R}\ud{j'}{j} \mca{R}\ud{k'}{k} \mca{R}\ud{l'}{l} \, \la_{i'j'k'l'}
\eeq
so its symmetry group is the set of matrices that leave $\la_{ijkl}$ invariant, that is to say
\beq
\msf{Stab}(\la) = \{ \, \mca{R} \in O(N) :\,  \la^{\mca{R}} = \la\, \} \subset O(N).
\eeq
For $N=1$, the Wilson-Fisher model~\reef{eq:Ising} is a one-loop fixed point with the smallest possible symmetry group, namely $\mbb{Z}_2$. Whether any CFTs with $N \geq 2$ share this property is an interesting open problem~\cite{Rychkov:2018vya}.

\subsection{Orbits}
\label{sec:orbits}

Two Lagrangians related by a field redefinition $\phi^i \mapsto \mca{R}\ud{i}{j} \phi^j$ are physically equivalent. However, they have different-looking couplings. For concreteness, consider two copies of the Wilson-Fisher fixed point:
\beq
\label{eq:twoIsing}
V(\phi^i) = \frac{1}{3} (\phi_1^4 + \phi_2^4).
\eeq
Now rotate the two fields by an angle $\theta$. This gives rise to a one-dimensional family of potentials
\beq
V(\phi^i) \mapsto V'(\phi^i) = \frac{3+\cos(4\theta)}{12}\, (\phi_1^4 + \phi_2^4) + \frac{\sin(4\theta)}{3}\, (\phi_1^3 \phi_2 - \phi_1 \phi_2^2) + \frac{1-\cos(4\theta)}{2}\,  \phi_1^2 \phi_2^2
\eeq
all of which describe the same physics. For a general coupling $\lambda_{ijkl}$, the set of equivalent couplings is known as its {orbit}:
\beq
\msf{Orb}(\la) = \{ \,\la^\mca{R} \; | \; \mca{R} \in O(N)\, \}.
\eeq
The ``size'' of the orbit of a tensor $\la_{ijkl}$ is closely related to its symmetry group: a version of the orbit-stabilizer theorem shows that
\beq
\label{eq:orbf}
\text{dim}\, \msf{Orb}(\la) = \text{dim}\,  O(N) - \text{dim}\,  \msf{Stab}(\la).
\eeq
For instance, the theory~\reef{eq:twoIsing} has a discrete symmetry group, so the previous formula correctly predicts that its orbit is one-dimensional.

As an important consequence, individual components of the coupling ($\la_{1111}$, $\la_{1112}$, $\ldots$) don't have a physical meaning by themselves: observables like critical exponents should not depend on a choice of frame, but instead they should be constant on orbits. It is therefore convenient to discuss \emph{invariants} of a given coupling, like $\norm{\la}^2 = \la_{ijkl}^2$. This idea will be explored in more detail in Section~\ref{sec:boundsMAIN}.

\subsection{Role of isotropy}
\label{sec:iso}

In much of the RG literature, fixed points are studied based on their putative symmetry group. That is to say that a global symmetry group $G \subset O(N)$ is chosen, and fixed points are then sought inside the subset of couplings which are invariant under $G$. If $G$ is sufficiently ``large'', the number of invariant couplings $I_4$ will be of order unity, so it is relatively straightforward to construct fixed points. For instance, the hypercubic symmetry group $\mbb{Z}_2^N \rtimes \mca{S}_N$ has $I_4 = 2$, hence the fixed point~\reef{eq:cubAns} could be found by solving for two coefficients.

Even though~\reef{eq:genLag} contained only a kinetic term and a quartic coupling, the full $\phi^4$ Lagrangian also contains mass terms, i.e.\@ couplings of the form $m_{ij}  \phi^i \phi^j$. Such terms are not generated in minimal subtraction, but they appear in more general renormalization schemes, and they definitely must be considered when doing RG computations directly in three dimensions. Enforcing invariance under some symmetry group $G$ also limits the number $I_2$ of such mass terms.

Much of the literature has focused on QFTs with $I_2 = 1$, said to be \emph{isotropic}. From a group theory point of view, this means investigating subgroups of $O(N)$ that have a unique quadratic invariant, $\dd_{ij}$. When $N$ is small, it is possible to classify such groups and to find all possible CFTs in the corresponding universality classes. This work has been done  for $N=2,3$ in~\cite{Zia:1974nv} and for $N=4$ in~\cite{Brezin:1985br}. For $N=2,3$, the only isotropic fixed points turn out to be the $O(N)$ CFT and $N$ copies of the Ising model; for $N = 4$ the solution is much richer.

In the rest of this work, we will \emph{not} impose isotropy, and instead look for completely general fixed points. Throughout the next section, we will however comment on the role of isotropy.

%% file: sections/bounds.tex

The number of couplings $\la_{ijkl}$ in a theory with $N$ fields is
\beq
\label{eq:dn}
d_N = \binom{N+3}{4} = \frac{1}{24} N(N+1)(N+2)(N+3)
\eeq
which grows rapidly with $N$. For $N=1$ there is a single coupling $\la_{1111}$, and the only non-trivial fixed point is the Wilson-Fisher CFT. For $N=2$ there are $d_2 = 5$ couplings, and it's already an interesting problem to classify all solutions~\cite{Osborn:2017ucf}. For $N =3,4,5,\ldots$ there are $d_N = 15, 35, 70, \ldots$ couplings, and no analytic methods are available to produce such a classification.

Instead, we can use a bottom-up method to prove theorems about the space of solutions. Assuming that $\la_{ijkl}$ is a real zero of the one-loop beta function~\reef{eq:oneloopbeta} for a given number of fields $N$, what can we say about $\la_{ijkl}$? Both in the older literature~\cite{Brezin:1973jt} and in more recent papers~\cite{Osborn:2017ucf,Rychkov:2018vya} this question was explored, and in this section we will pursue this strategy further.

\subsection{Bounds on $\norm{\la}$}
\label{bounds:real}

Let us start by reviewing one of the main results of Ref.~\cite{Rychkov:2018vya} by Rychkov and Stergiou. They found that if $\la_{ijkl}$ is a one-loop fixed point, then its norm $\norm{\la}^2 = \la_{ijkl}^2$ obeys the following inequality:
\beq
\label{eq:RSbound}
 \norm{\la}^2 \leq
\begin{cases}
  \frac{1}{36}(3+4\sqrt{2}) \approx 0.240468 \quad  & N=2\\
  \frac{1}{12}(1+2 \sqrt{3}) \approx 0.372008       & N=3\\
  \frac{1}{8}N                                      & N \geq 4
\end{cases}\;.
\eeq
The proof of Eq.~\reef{eq:RSbound} relies on a clever application of the Cauchy-Schwarz inequality, and we refer to~\cite{Rychkov:2018vya} for details.

We will now present a new result, namely that the norm of any \emph{interacting} fixed point $\la_{ijkl}$ is bounded from below as well:
\beq
\label{eq:WFb}
\norm{\la} \geq \tfrac{1}{3}.
\eeq
Of course, $\la_{ijkl} = 0$ is also a fixed point --- it describes $N$ decoupled free scalars. To show~\reef{eq:WFb}, it will be useful to define an operator $\vee$ on the space of tensors. If $A_{ijkl}$ and $B_{m_1 \dotsm m_r}$ are two symmetric tensors of rank $4$ resp.\@ $r \geq 2$, then let\footnote{For the $r=4$ case, this operator appeared already in~\cite{Michel:1985pq}.}
\beq
\label{eq:veeprod}
(A \vee B)_{i_1 \dotsm i_r} \ldef A_{mn(i_1 i_2} B_{i_3 \dotsm i_r) m n}
\eeq
where $X_{(i_1 \dotsm i_r)}$ denotes symmetrization over the indices $\{i_k\}$. For instance, for $r=4$ this $\vee$ product is given by
\begin{multline}
(A \vee B)_{ijkl} = \tfrac{1}{6} \big( A_{ijmn} B_{klmn} + A_{ikmn} B_{jlmn} + A_{ilmn} B_{jkmn} \\ + A_{jkmn} B_{ilmn} + A_{jlmn} B_{ikmn} + A_{klmn} B_{ijmn} \big).
\end{multline}
In particular, the one-loop $\beta$ function~\reef{eq:oneloopbeta} can be written as
\beq
\beta(\la)_{ijkl} = - \la_{ijkl} + 3(\la \vee \la)_{ijkl}
\eeq
omitting the superscript on $\beta^{(1)}$ from now on. Two key properties of the $\vee$ product are stated below.
\begin{fact}
  The $\vee$ product~\reef{eq:veeprod} satisfies
    \beq
    \label{eq:ab}
    \norm{A \vee B} \leq \norm{A} \norm{B}
    \eeq
    where $\norm{B}^2 = B_{i_1 \dotsm i_r}^2$ is the usual norm on tensors. If~\reef{eq:ab} is saturated, then there exist constants $c,c'$ and a unit vector $u_i$ such that
    \beq
    \label{eq:saturated}
    A_{ijkl} = c \, u_i u_j u_k u_l,
    \quad
    B_{i_1 \dotsm i_r} = c'\, u_{i_1} \dotsm u_{i_r}.
    \eeq
 \end{fact}
\noindent The proof of~\reef{eq:ab} and~\reef{eq:saturated} is given in Appendix~\ref{AppB}.

Coming back to the $\beta$ function, we can use the above inequality to get
\beq
\label{eq:lowerbd}
\norm{\la} =3 \norm{\la \vee \la} \leq  3 \norm{\la}^2
\quad
\Rightarrow
\quad
{
  \la = 0
  \quad
  \text{or}
  \quad
  \norm{\la} \geq \tfrac{1}{3}
}
\eeq
as promised. When the inequality~\reef{eq:lowerbd} is saturated, Eq.~\reef{eq:saturated} implies that $\la$ factorizes as follows:
\beq
\label{eq:isingDef}
|\la \vee \la| = |\la|^2 \quad
\Rightarrow
\quad
\la_{ijkl} = c\, u_i u_j u_k u_l\,.
\eeq
The overall normalization $c$ is fixed to be $1/3$ by requiring that $\la_{ijkl}$ is a fixed point. In conclusion, any fixed point obeying $\norm{\la} = 1/3$ is equal the {Wilson-Fisher} CFT~\reef{eq:Ising}.

\subsection{Bounds on $O(N)$ invariants of $\la$}
\label{sec:realinv}

We can move ahead and further restrict the space of fixed points, by carefully analyzing the solutions of $\beta(\la) = 0$. As announced before, it will be useful to consider $O(N)$ invariants of the coupling $\la_{ijkl}$, since such invariants don't depend on a choice of frame.

To define such invariants, we recall that symmetric tensors are reducible under $O(N)$. If $V_r$ denotes the vector space of rank-$r$ symmetric tensors, then we have the following decomposition in terms of irreducible representations:
\beq
\label{eq:tensordec}
V_r = [r] \oplus [r-2] \oplus \dotsm \oplus [r \text{ mod } 2]
\eeq
writing $[\l]$ for the traceless symmetric rank-$\l$ irrep of $O(N)$. 
Let us make the decomposition~\reef{eq:tensordec} explicit for the $r=4$ case. For an arbitrary tensor $A_{ijkl}$ the decomposition in invariant components is determined by projectors $\mca{P}_\l : V_4 \to [\l] \subset V_4$ as follows:
\bsub
\label{eq:projectors}
\begin{align}
  (\mca{P}_0 A)_{ijkl} &= \frac{3}{N(N+2)}\, \dd_{(ij}\dd_{kl)} A_{mmnn}\\
  (\mca{P}_2 A)_{ijkl} &= \frac{6}{N+4}\,  \dd_{(ij} A_{kl) mm} -  \frac{2(N+2)}{N+4}\, (\mca{P}_0 A)_{ijkl}\\
  (\mca{P}_4 A)_{ijkl} &= A_{ijkl} -  (\mca{P}_2 A)_{ijkl} -  (\mca{P}_0 A)_{ijkl}.
\end{align}
\esub
We note that the projector $\mca{P}_2$ has a simple physical interpretation. In an \emph{isotropic} quantum field theory, the coupling $\la_{ijkl}$ must obey
\beq
\label{eq:tracecond}
\la_{ijkk} = z \cdot \dd_{ij}
\eeq
for some $z \in \mbb{R}$ --- after all, $\dd_{ij}$ is the only allowed invariant. A tensor satisfying~\eqref{eq:tracecond} then necessarily has $\mca{P}_2 \lambda = 0$.

To proceed, we will can define $O(N)$ invariants $\ba_\ell(\la)$ by means of the projectors~\reef{eq:projectors}. There will be one linear invariant $\ba_0(\la)$ and two quadratic invariants $\ba_{2,4}(\la)$ which by construction obey
\beq
\ba_\l(\mca{P}_\l \la) = \ba_\l(\la)
\qaq
\ba_{\l}(\mca{P}_{\l'} \la) = 0
\quad
\text{if}
\quad
\l \neq \l'.
\eeq
Explicitly, they are given by
\beq
\label{eq:invDef}
\ba_0(\la) \ldef \la_{iijj},
\quad
\ba_2(\la) \ldef \la_{ijkl} (\mca{P}_2 \la)_{ijkl},
\quad
\ba_4(\la) \ldef \la_{ijkl} (\mca{P}_4 \la)_{ijkl}.
\eeq
Notice that the quantity $\ba_0(\la)$ can have either sign (since $\ba_0(\la)$ is odd under $\la \mapsto -\la$) while $\ba_2(\la)$ and $\ba_4(\la)$ are positive, These invariants are related to $|\la|^2$ via the equation
\beq \label{eq:invrel}
\norm{\la}^2 = \frac{3}{N(N+2)}\, \ba_0(\la)^2 +  \ba_2(\la) +  \ba_4(\la)
\eeq
which can for example be used to eliminate $\ba_4(\la)$. Following the previous discussion, $\ba_2(\la)$ always vanishes in isotropic theories.

For concreteness, let us compute the above invariants for the Wilson-Fisher CFT tensored with $N-1$ free fields~\reef{eq:Ising}. We get
\beq
\label{eq:ising-inv}
\la = \la^\text{WF}:
\quad
\norm{\la}^2 = \frac{1}{9},
\quad
\ba_0(\la) = \frac{1}{3},
\quad
\ba_2(\la) = \frac{2(N-1)}{3N(N+4)}.
\eeq
Next, we can consider the $O(N)$ fixed point~\reef{eq:ONfp}
\bsub
\beq
\label{eq:ON-def}
\la^{O(N)}_{ijkl} = \frac{3}{N+8} \, \dd_{(ij} \dd_{kl)}
\eeq
which obeys
\beq
\label{eq:ON-inv}
\norm{\la}^2 = \frac{3N(N+2)}{(N+8)^2},
\quad
\ba_0(\la) = \frac{N(N+2)}{N+8},
\quad
\ba_2(\la) = 0.
\eeq
\esub
In the tables below, we have collected a list of known fixed points of $N=3$ and $N=4$ fixed points from the literature (see for instance Ref.~\cite{Brezin:1985br}, and in particular Ref.~\cite{Codello:2020lta} for three new fixed points) and computed their invariants. Notice that most theories in these tables are \emph{decoupled}, i.e.\@ they are tensor products of fixed points with fewer than $N$ scalars. Among the theories listed below, only three CFTs with $N=3$ fields and five CFTs with $N=4$ fields are fully interacting. 

\begin{center}
\begin{tabular}{ |p{4.5cm}||p{2.cm}|p{1.8cm}|p{1.8cm}|p{1.8cm}|  }
 \hline
 \multicolumn{5}{|c|}{Table of invariants for $N=3$ fixed points} \\
 \hline
  & $ |\lambda|^2$  & $\ba_0$ & $\ba_2$ & $\ba_4$ \\
 \hline
Gaussian   & 0    &0& 0 & 0 \\
Ising &1/9 & 1/3 &  4/63& 8/315 \\
Ising$^2$ &2/9 & 2/3 &  4/63 & 22/315  \\
Ising$^3$ &1/3 & 1 &  0& 2/15  \\
O(2)&   6/25  &4/5  & 16/175  & 18/875  \\
O(2) + Ising & 79/225 & 17/15 &  4/1575 & 722/7875  \\
Cubic& 10/27 & 4/3 &  0 & 2/135  \\
Biconical & 0.370451 & 1.33713 & 0.0002184 & 0.012651  \\
 O(3) & 45/121 &15/11 & 0 & 0 \\
\hline
\end{tabular}
\end{center}

\begin{center}
\begin{tabular}{ |p{4.5cm}||p{2.cm}|p{1.8cm}|p{1.8cm}|p{1.8cm}|  }
 \hline
 \multicolumn{5}{|c|}{Table of invariants for $N=4$ fixed points} \\
 \hline
  & $ |\lambda|^2$  & $\ba_0$ & $\ba_2$ & $\ba_4$ \\
 \hline
Gaussian   & 0    &0& 0 & 0 \\
O(2)&   6/25  &4/5  & 3/25  &1/25  \\
 O(3) & 45/121 &15/11 & 225/1936 & 45/1936  \\
 O(4)    & 1/2 & 2 & 0 & 0 \\
O(2) + O(2) &12/25 & 8/5 &  0& 4/25  \\
 Ising &1/9 & 1/3 &  1/16& 5/144 \\
Ising$^2$ &2/9 & 2/3 &  1/12& 1/12  \\
Ising$^3$ &1/3 & 1 &  1/16& 7/48  \\
Ising$^4$ &4/9 & 4/3 &  0& 2/9  \\
O(3) + Ising &526/1089 & 56/33 &  1/121& 125/441  \\
O(2) + Ising & 79/225 & 17/15 &  33/400 & 389/3600   \\
O(2) + 2 Ising& 104/225 & 22/15 &  1/300 & 19/100  \\
$N=3$ Cubic& 10/27 & 4/3 &  1/9 & 1/27  \\
$N=3$ Cubic + Ising & 13/27 & 5/3 &  1/144 & 7/432  \\
$S_1;S_2$ from di-pentagonal &220/441 & 40/21 &  0 & 20/441  \\
$N=3$ biconical &0.370451 & 1.33713  & 0.111935 & 0.0350273  \\
$N=3$ biconical + Ising & 0.481562 & 1.67046 & 0.0072945 & 0.125463  \\
$O(2)$ ($v_1$ from \cite{Codello:2020lta})  & 0.499606 & 1.95458 & 0.00020497 & 0.0218515  \\
 $D_4 \times \mathbb{Z}_2 $ ($v_2$ from \cite{Codello:2020lta}) & 0.499144 & 1.92641 & 0.00026922  & 0.0349939 \\
$S_3 \times \mathbb{Z}_2^2$ ($v_3$ from \cite{Codello:2020lta})  & 0.499115 & 1.92406 & 0.00024596 & 0.0361167 \\
\hline
\end{tabular}
\end{center}

In what follows, we will derive bounds on the invariants $\ba_{\l}(\la)$ for that are valid at any fixed point. This idea is not new: essentially, we are generalizing results due to Br\'{e}zin et al.\@ from Ref.~\cite{Brezin:1973jt}. The new ingredient in our work is that we allow for anisotropic theories; in their work, they only considered couplings of the form
\beq
\label{eq:isoans}
\la_{ijkl} = c \cdot (\dd_{ij} \dd_{kl} + \text{symm}) + W_{ijkl}
\eeq
where $W_{ijkl}$ is a traceless tensor. Couplings of the form~\reef{eq:isoans} automatically have $\ba_2(\la) = 0$. 

Going back to our computation, we start by applying $\ba_0$ to the fixed point condition $\beta(\la) = 0$, making use of the identity\footnote{A version of~\reef{eq:a0id} for tensors obeying $\mca{P}_2A = \mca{P}_2 B=0$ already appeared in~\cite{Michel:1985pq}.}
\beq
\label{eq:a0id}
3 \ba_0(A \vee B) = N^{-1} \ba_0(A)\ba_0(B) + \tfrac{1}{6}(N+4) A_{ijkl} (\mca{P}_2 B)_{ijkl}  + 2 A_{ijkl} B_{ijkl}
\eeq
which yields
\beq
\label{eq:quad}
\frac{1}{2N} \, \ba_0(\la)(N- \ba_0(\la))  =  \norm{\la}^2+ \frac{1}{12}(N+4) \ba_2(\la).
\eeq
This is an important relation, showing that the invariant $\ba_2(\la)$ is redundant.

Now, notice that the RHS of~\reef{eq:quad} is bounded from below by $1/9$, assuming that $\la \neq 0$. This restricts the range of $\ba_0$:
\beq
\label{eq:naivea0bd}
\la_{ijkl} = 0 \qor \left|\ba_0(\la) - \frac{N}{2}\right| <   \frac{N}{2} \sqrt{1-\frac{8}{9N}}\,.
\eeq
This bound is not yet optimal. To refine it, we can simply replace $|\la|^2$ in~\reef{eq:quad} by the RHS of~\reef{eq:invrel}, which leads to
\beq
\label{eq:quad2}
\frac{1}{2\varrho^+_N} \, \ba_0(\la)\!\left[ \varrho^+_N- \ba_0(\la)\right]  = \frac{1}{12}(N+16) \ba_2(\la) +   \ba_4(\la),
\quad
\varrho^+_N \ldef \frac{N(N+2)}{N+8}.
\eeq
Since the RHS of~\reef{eq:quad2} is positive, this immediately implies that
\beq
\label{eq:a0bd}
{
\la = 0
\quad
\text{or}
\quad
\varrho^{-}_N < \ba_0(\la) \leq \varrho^+_N\,,
\quad
\varrho^{-}_N \ldef \frac{N}{2}\!\left[1 -   \sqrt{1 - \frac{8}{9N} }\;\right].
}
\eeq
The upper bound is optimal, since $\varrho_N^+$ is precisely the value of $\ba_0$ of the $O(N)$ fixed point from~\reef{eq:ON-inv}.

We can also obtain an upper bound on $|\lambda|^2$, starting from the relation~\reef{eq:quad}. Since $\ba_2(\la) \geq 0$, it follows that
\beq
|\lambda|^2 \leq \frac{1}{2N} \, \ba_0(\la)(N -\ba_0(\la)).
\eeq
As a function of $\ba_0(\la)$ the RHS attains a maximum at $\ba_0(\la) = \th N$, so in particular $|\la|^2 \leq \tfrac{1}{8} N$, which is the previously announced result~\reef{eq:RSbound}. However, for low values of $N$, the maximum $ \th N$ is larger than $\varrho_N^+$.  This is the case for $N=2,3$, and consequently for these values of $N$, $\norm{\la}^2$ is maximal at the $O(N)$ fixed point. In summary, we find that
\beq
\label{eq:RSref}
\norm{\la}^2 \leq \begin{cases} \norm{\la^{O(N)}}^2 = \frac{6}{25},\, \frac{45}{121} &\text{for} \quad N=2,3\\
  \frac{1}{8} N &\text{for} \quad N \geq 4
  \end{cases}\;.
\eeq
For $N=2$ and $N=3$ this is a slight improvement compared to Eq.~\reef{eq:RSbound}: for instance $45/121 \approx 0.3719$ is slightly smaller than $0.3720$.

We already proved that $\norm{\la} \geq \tfrac{1}{3}$ for any interacting fixed point. This lower bound can be refined using the condition $\ba_4(\la) \geq 0$. Using~\reef{eq:invrel} and~\reef{eq:quad}, we find
\bsub
\begin{align}
  0 \leq \ba_4(\la) &= \norm{\la}^2 - \ba_2(\la) - \frac{3}{N(N+2)}\, \ba_0(\la)^2\\
  &= \frac{1}{N+4} \left[ (N+16) \norm{\la}^2 - 6\ba_0(\la) + \frac{3}{N+2} \, \ba_0(\la)^2 \right]
\end{align}
\esub
and therefore
\beq
\label{eq:tomax}
\norm{\la}^2 \geq \max\left\{\frac{1}{9},\, \frac{3\ba_0(\la)}{(N+2)(N+16)} \left[2N+4 - \ba_0(\la)\right]\right\}.
\eeq
The two functions appearing on the RHS are equal if
\beq
\ba_0(\la) = N+2- \sqrt{\tfrac{2}{27}(N+2)(13N+19)} \rdef \varrho^\sharp_N\, .
\eeq
For $\varrho^{-}_N < \ba_0(\la) \leq \varrho^\sharp_N$, the norm $\norm{\la}^2$ is bounded from below by $\tfrac{1}{9}$, whereas for $\varrho^\sharp_N < \ba_0(\la) \leq \varrho^{+}_N$ the relevant bound is given by the non-trivial function of $\ba_0(\la)$ appearing in~\reef{eq:tomax}.

\begin{figure}[htb]
    \begin{center}
    \hspace{0mm}
    \includegraphics[scale=.56]{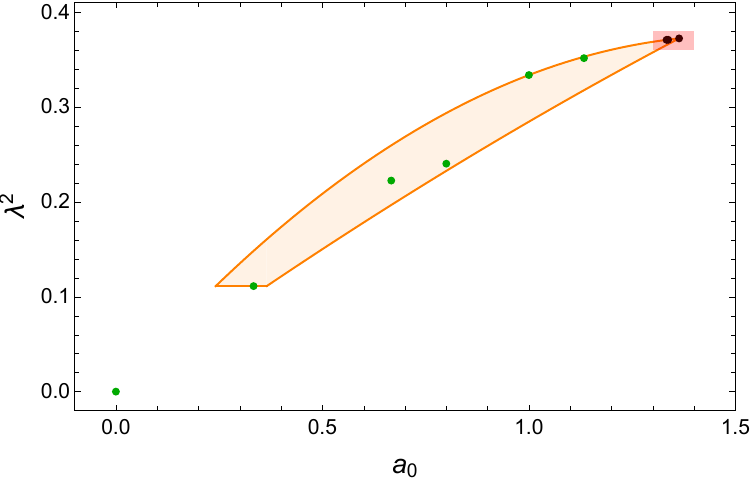}
    \hspace{2mm}
    \includegraphics[scale=.42]{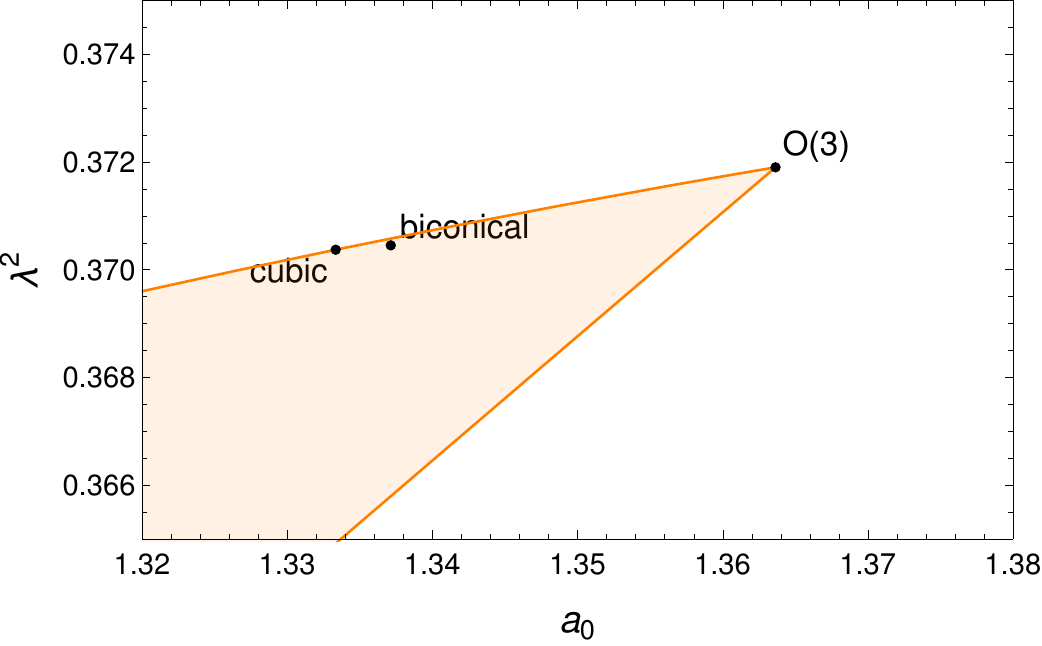}
    \caption{\label{fig:plota0squaredn3} Left: allowed region (orange) in the $\ba_0, |\lambda|^2$ plane for theories with $N=3$ fields. Dots represent known fixed points, as reported in the previous table. Right: zoom near the $O(3)$ fixed point. In both plots, decoupled fixed points are shown in green, fully interacting fixed points in black.}
  \end{center}
   \end{figure}

\begin{figure}[htb]
    \begin{center}
    \hspace{0mm}
    \includegraphics[scale=.55]{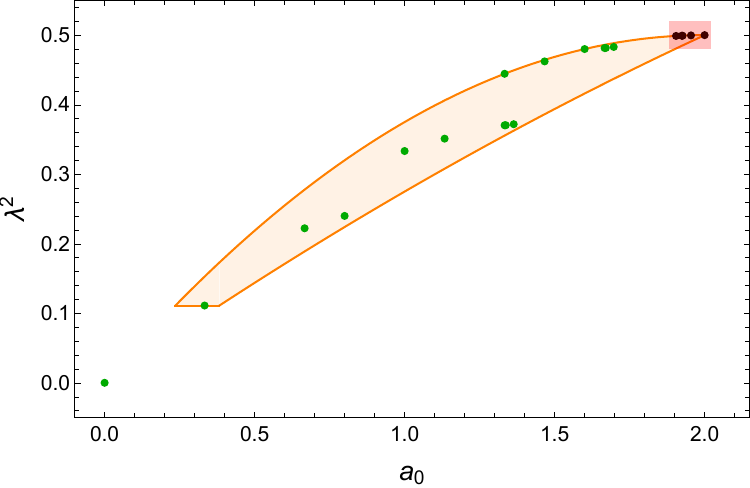}
    \hspace{2mm}
    \includegraphics[scale=.42]{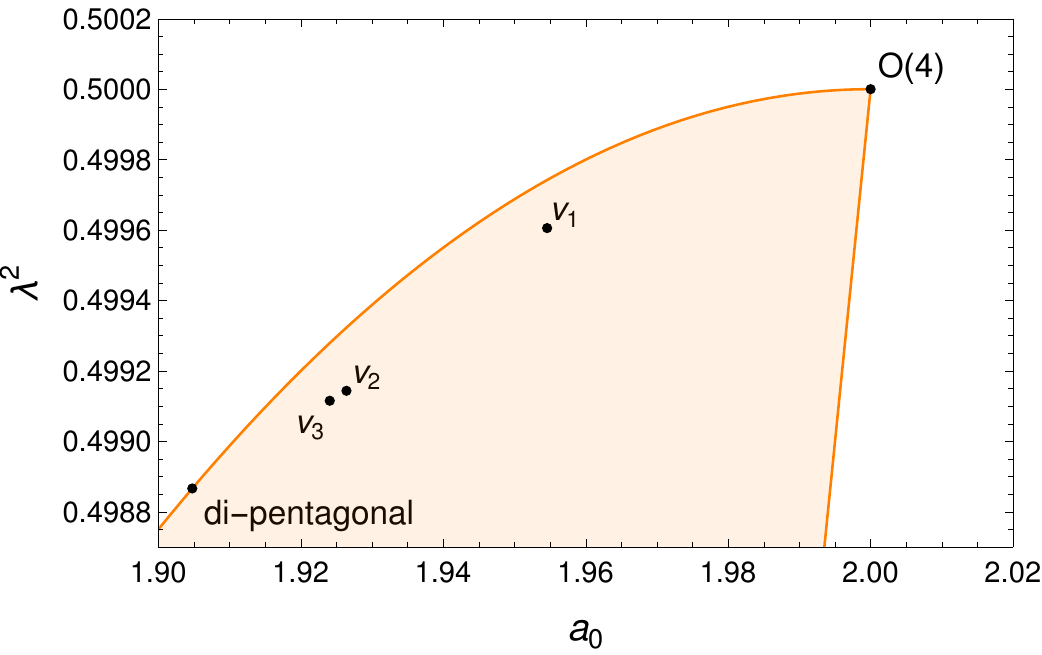}
    \caption{\label{fig:plota0squaredn4} Left: allowed region (orange) in the $\ba_0, |\lambda|^2$ plane for $N=4$. Dots represent known fixed points. Right: zoom near the $O(4)$ fixed point. Dots are color-coded as in Fig.~\ref{fig:plota0squaredn3}.}
  \end{center}
  \end{figure}

For $N=3$ and $N=4$, the allowed regions in the plane spanned by $\ba_0(\la)$ and $\norm{\la}^2$ are shown in Figures~\ref{fig:plota0squaredn3} and~\ref{fig:plota0squaredn4}, along with the location of all known fixed points. 

\begin{figure}[htb]
    \begin{center}
    \hspace{0mm}
    \includegraphics[scale=.54]{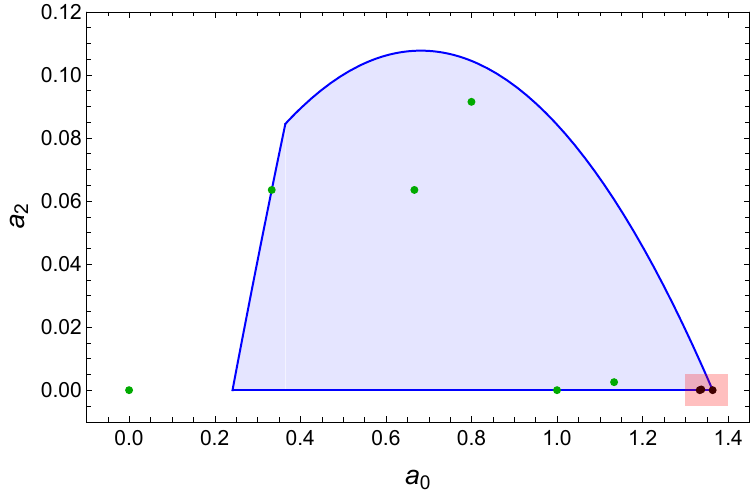}
    \hspace{2mm}
    \includegraphics[scale=.42]{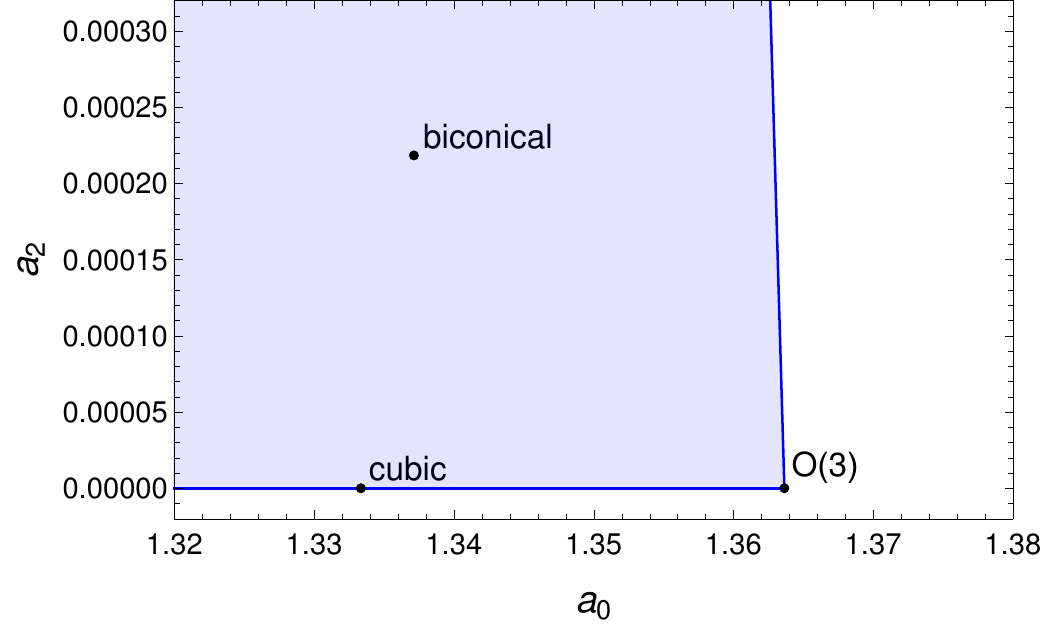}
    \caption{\label{fig:plota0a2n3} Left: allowed region (blue) in the $\ba_0, \ba_2$ plane for $N=3$. Dots represent known fixed points. Right: zoom near the $O(3)$ fixed point. Dots are color-coded as in Fig.~\ref{fig:plota0squaredn3}.}
  \end{center}
 \end{figure}

\begin{figure}[htb]
    \begin{center}
    \hspace{0mm}
    \includegraphics[scale=.56]{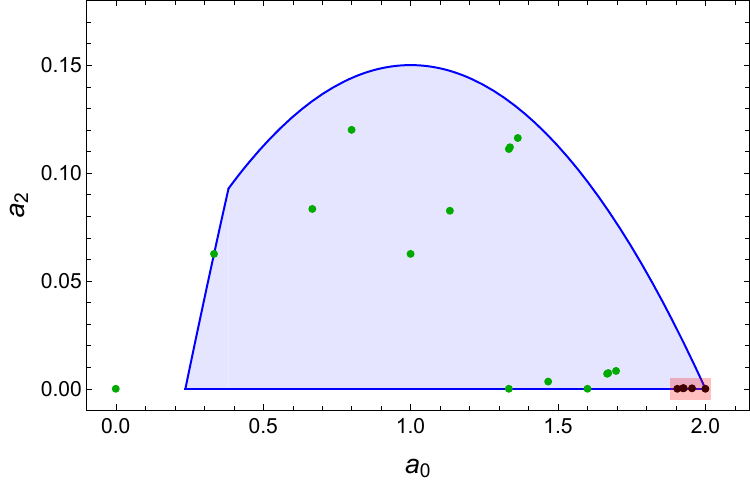}
    \hspace{2mm}
    \includegraphics[scale=.42]{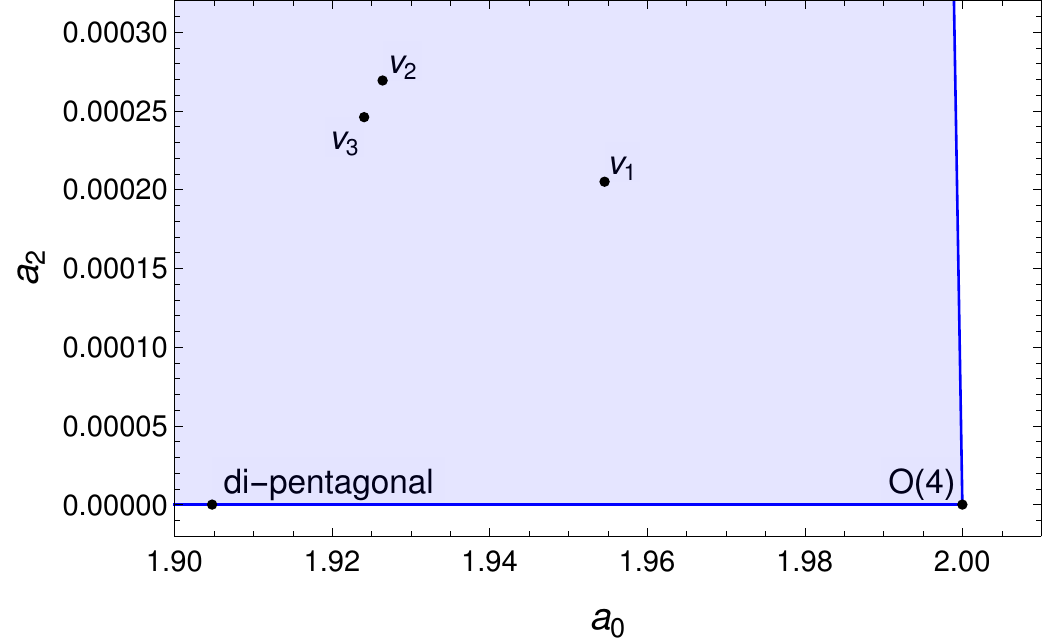}
    \caption{\label{fig:plota0a2n4} Left: allowed region (blue) in the $\ba_0, \ba_2$ plane for $N=4$. Dots represent known fixed points. Right: zoom near the $O(4)$ fixed point. Dots are color-coded as in Fig.~\ref{fig:plota0squaredn3}.}
  \end{center}
 \end{figure}

  Finally, we can turn to the anisotropy $\ba_2(\la)$, which by construction is non-negative. To obtain an upper bound, we start by rewriting Eq.~\reef{eq:quad}:
  \beq
  \label{ba2}
  0 \leq \frac{1}{12}(N+4)\ba_2(\la) \leq  \frac{1}{2N} \, \ba_0(\la)(N-\ba_0(\la)) - \norm{\la}^2.
  \eeq
  Depending on whether $\ba_0(\la)$ is smaller or larger than $\rho^\#_N$, the RHS can then be bounded by~\reef{eq:tomax}.
For small $\ba_0(\la)$, we get the inequality
\beq
\ba_0(\la) \leq \varrho_N^\sharp
\quad
\Rightarrow
\quad
0 \leq \ba_2(\la) \leq \frac{6}{N(N+4)} \left[\ba_0(\la)(N-\ba(\la)) - \frac{2}{9}N \right].
\eeq
Meanwhile, for larger $\ba_0(\la)$ we find
\beq
\ba_0(\la) \geq \varrho_N^\sharp
\quad
\Rightarrow
\quad
0 \leq \ba_2(\la) \leq \frac{6}{(N+16)\varrho_N^{+}} \, \ba_0(\la)(\varrho_N^{+}-\ba_0(\la)).
\eeq
This allowed region in the plane parametrized by $\ba_2(\la)$ and $\norm{\la}^2$ is plotted for $N=3,4$ in Figures~\ref{fig:plota0a2n3} and~\ref{fig:plota0a2n4}.
 
Scanning over all allowed values of $\ba_0(\la)$, we conclude in particular that 
\beq
\ba_2(\la) \leq  \frac{3\varrho_N^+}{2(N+16)} < \frac{3}{2}\,.
\eeq
In physics terms, this means that the anisotropy of a coupling $\la_{ijkl}$ is at most of order unity, although naively one could expect that there are fixed points with $\ba_2(\la)$ of order $\norm{\la}^2 \sim \frac{1}{8} N$. 

The rightmost parts of figures~\ref{fig:plota0squaredn3},~\ref{fig:plota0squaredn4},~\ref{fig:plota0a2n3} and~\ref{fig:plota0a2n4} show zoomed-in plots of theory space near the $O(3)$ and $O(4)$ fixed points, which live at kinks of the allowed region $\mca{D}$. Especially for $N=4$, we find that there are many fixed points that extremely close to the $O(4)$ theory. In particular, the three new $N=4$ fixed points found in~\cite{Codello:2020lta} --- labeled by $v_1$, $v_2$ and $v_3$ --- have this property. These new fixed points  also have $\ba_2(v_i) \neq 0$ but of the order of $10^{-4} \ll 1$.  
It would be interesting to find a deeper explanation why there are so many fixed points near these $O(N)$ kinks.

Finally, let us compare our results to those of Ref.~\cite{Brezin:1973jt} by Br\'{e}zin et al. In that seminal work, a version of~\reef{eq:quad} was found with $\ba_2(\la) = 0$, along with a counterpart of~\reef{eq:quad2}. The authors of~\cite{Brezin:1973jt} correctly found that for isotropic theories $\ba_0(\la) \leq \varrho_N^+$, and later Michel~\cite{Michel:1985pq} pointed out that for $N<4$, the $O(N)$ model maximizes $\norm{\la}^2$, again assuming isotropy. Interestingly, both of these results are ultimately independent of the isotropy assumption.

%% file: sections/anomalous.tex

So far, we derived bounds on the couplings $\la_{ijkl}$ describing a general CFT at order $\vareps$ in the epsilon expansion. In what follows, we will instead focus on the the \emph{spectrum} of such a CFT, that is to say its one-loop anomalous dimensions. We will start by briefly reviewing the problem of computing one-loop anomalous dimensions and proceed to derive explicit bounds for operators without derivatives.

\subsection{One-loop anomalous dimensions}

To start, we briefly review the computation of anomalous dimensions in perturbative quantum field theory. Local operators in the \emph{free} scalar CFT are of the form
\beq
\label{eq:Ofree}
\text{free theory}:
\quad
\mca{O}^{i_1 \dotsm i_n} = \prod_{i=1}^n \pd^{\ell_i} \phi^{i_1}
\eeq
where $\pd^{\ell}$ is shorthand for a product of $\ell$ derivatives, say $\pd^{\ell} = \pd_{\mu_1} \dotsm \pd_{\mu_\ell}$. Such operators can be organized into irreducible representations of the rotation group $O(d)$ and of the global symmetry group, $O(N)$. In $d=4-\vareps$ dimensions, the operator~\reef{eq:Ofree} has scaling dimension
\beq
\DD_{\text{free}} = n \cdot \DD_\phi + \sum_i \ell_i = n (1- \th \vareps) + \sum_i \ell_i
\eeq
since $\DD_\phi = \th(d-1) = 1 - \th \vareps$ is the dimension of a free scalar field.

If the theory is instead deformed by a coupling $\la_{ijkl}$ such that an IR fixed point is reached, the scaling dimensions change. There will be a {new} set of scaling operators $[\Oo_\a]$ with non-zero anomalous dimensions, so
\beq
\DD_\a = \DD_{\a,\text{free}} + \gamma_\a^{(1)} \cdot \vareps +  \gamma_\a^{(2)} \cdot \vareps^2 + \ldots
\eeq
where the $\gamma_\a^{(n)}$ are coefficients that can be computed using an $n$-loop computation in perturbation theory. The scaling operators $[\Oo_\a]$ can still be organized into definite representations of $O(d)$; however, they are now labeled by representations of the symmetry group of $\la_{ijkl}$, which may be smaller than $O(N)$.

The scaling operators $[\Oo_\a]$ can be expressed as linear combinations of free theory operators of the form~\reef{eq:Ofree}. The computation of the relevant change-of-basis matrix at general orders in perturbation theory is complicated, and we will not discuss in this work. However, it is known that expressions for scaling operators \emph{at one loop} can be computed by solving a diagonalization problem that only involves free-theory CFT data~\cite{jc}. To be precise, let
\beq
\mca{V} = \NO{\la_{ijkl} \, \phi^i \phi^j \phi^k \phi^l}
\eeq
be the operator that perturbs the Gaussian theory, and let $\Oo_\rho$ be a generic operator in the free theory, say a Lorentz scalar, of 4$d$ scaling dimension $\DD_\text{free} = r$, which is necessarily an integer, $r \in \{1,2,3,\ldots\}$. 
In the 4$d$ Gaussian CFT, we have the following OPE:
\begin{multline}
\label{eq:VO}
\mca{V}(x) \Oo_\rho(0) = \frac{1}{|x|^4} \sum_\sigma  {\sf C}\ud{\sigma}{\rho}(\la) \Oo_\sigma(0) + \text{non-scalar operators} \\ + \text{scalar operators of 4$d$ dimension $\neq r$} 
\end{multline}
where ${\sf C}\ud{\sigma}{\rho}$ is a matrix of OPE coefficients that can e.g.\@ be determined using Wick contractions. It can be shown that scaling operators of the IR CFT are precisely of the form $[\Oo_v] = v^\rho \, \Oo_\rho$, where $v^\rho$ is an eigenvector of the matrix ${\sf C}$:
\beq
\label{eq:mixing}
{\sf C}\ud{\sigma}{\rho}(\la)  v^\rho = \omega(v) v^\sigma.
\eeq
Moreover, the one-loop anomalous dimension $\ga_{v}^{(1)}$ of the operator $[\Oo_v]$ is proportional to $\omega(v)$:\footnote{The constant of proportionality depends on the normalization of the field $\phi$. From here on out, we use the normalization $\expec{\phi(x)\phi(0)} = 1/|x|^{2}$.}
\beq
\label{eq:mix2}
\ga_{v}^{(1)} = \frac{1}{12}\,  \omega(v).
\eeq
Notice that the eigenvalue problem~\reef{eq:mixing} only involves operators of 4$d$ dimension $r$, even though the OPE $\mca{V} \times \Oo_\rho$ also contains operators with dimensions other than $r$.

The formulas~\reef{eq:VO},~\reef{eq:mixing} and~\reef{eq:mix2} provide an explicit recipe to compute one-loop anomalous dimensions of scalar operators of dimension $r$. It is perhaps illustrative to spell this out for the case $r=2$. All operators with 4$d$ dimension $\DD_\text{free} = 2$ are of the form $\Oo_{}^{ij} = \NO{\phi^i \phi^j}$ or $\pd_\mu \phi^i$, but the latter have have non-trivial Lorentz spin so they don't mix with the $\Oo_{}^{ij}$. Using Wick's theorem, we have the following OPE:
\beq
\mca{V}(x) \Oo^{ij}(0)= \frac{12}{|x|^4} \, \la_{ijkl} \, \Oo_{}^{kl}(0) + \ldots
\eeq
omitting operators with dimension $\DD_\text{free} \neq 2$ and spinning operators. In the case of the Wilson-Fisher CFT with $N=1$ we have $\la_{1111} = 1/3$, and there is a unique operator $\Oo_{}^{11} = \NO{\phi^2}$, so~\reef{eq:mix2} predicts that the renormalized operator $[\phi^2]$ has scaling dimension
\beq
\DD_{\phi^2} = 2 - \vareps + 12 \cdot \frac{1}{3} \cdot \frac{1}{12} \, \vareps + \sO(\vareps^2)
\quad
\Rightarrow
\quad
\ga_{\phi^2}^{(1)} = \frac{1}{3}.
\eeq
This result is well-known and can be obtained using other means, e.g.~\cite{Peskin:1995ev}. It is straightforward to generalize to $N \geq 2$ fields.

For operators of classical dimension $r \geq 4$, one might expect mixing of operators of the form $\NO{\phi^r}$ (without derivatives) with composite operators built out of $m$ derivatives and $r- m$ fields, say $\NO{\phi^{r-4} \pd_\mu \phi \pd^\mu \phi}$ having $m=2$. At one-loop order, it can be shown~\cite{Hogervorst:2015akt} that such mixing does not arise, based on a careful analysis of the matrix $\msf{C}$ appearing in Eq.~\reef{eq:VO}. It is therefore possible to study operators built out of $r$ fields $\phi^i$ \emph{without gradients} by a very modest generalization of the $r=2$ diagonalization problem treated above. If we let
\beq
\Oo^{i_1 \dotsm i_r} = \NO{\phi^{i_1} \dotsm \phi^{i_r}}
\eeq
then
\beq
\label{eq:spinningGen}
\mca{V}(x) \Oo_{i_1 \dotsm i_r}(0) = \frac{6 r(r-1)}{|x|^4} \la_{k l(i_1 i_2} \Oo_{i_3 \dotsm i_r) k l}(0)  + \ldots.
\eeq
As a consequence, if $v = v_{i_1 \dotsm i_r}$ solves the eigenvalue problem
\beq
\label{eq:genEV}
{
\frac{B}{16\pi^2}\;  \frac{r(r-1)}{2} \; \la \vee v = \gamma_v^{(1)} \, v
}
\eeq
then $[\Oo_{v}] = v_{i_1 \dotsm i_r} \Oo^{i_1 \dotsm i_r}$ is a scaling operator at the IR fixed point with one-loop anomalous dimension $\ga_v^{(1)}$. In~\reef{eq:genEV} we have allowed for a general normalization of the couplings, to facilitate comparison to the literature.\footnote{Versions of equation~\reef{eq:genEV} have appeared in the recent literature, for instance in~\cite{Antipin:2019vdg,Codello:2019vtg}.}

Notice that for fixed $r$, the number of such operators is equal to the number of symmetric tensors of rank $r$, that is to say
 \beq
 d_{N,r} = \binom{N+r-1}{r}
 \eeq
which behaves as $N^r/r!$ at large $N$.  For a given coupling $\la_{ijkl}$, finding anomalous dimensions $\ga_{v}^{(1)}$ for rank-$r$ operators is therefore equal to diagonalizing a matrix of size $d_{N,r} \times d_{N,r}$. 

The cases $r=0,1$ and $r=3,4$ deserve special treatment. The unique operator with $r=0$ is the identity operator, which does not get renormalized. The fundamental fields $\phi^i$ only get renormalized at order $\vareps^2$, by a two-loop diagram. This explains the prefactor $r(r-1)$ in the eigenvalue equation~\reef{eq:genEV}. For $r=3$, there are $N$ operators of the form
\beq
\Oo_i = \la_{ijkl} \, \NO{ \phi^j \phi^k \phi^l },
\qquad
i=1,\ldots,N
\eeq
which have anomalous dimension $\ga_{\Oo_i}^{(1)} = 1$. These operators therefore satisfy
\beq
\DD_{\Oo_i} = 3(1-\th \vareps)  + 1 \cdot \vareps + \sO(\vareps^2) = \DD_\phi + 2 + \sO(\vareps^2)
\eeq
as required because of multiplet recombination: in the IR theory, the field $\phi^i$ combines with $\Oo_i$ to form a long multiplet~\cite{Rychkov:2015naa}. The perturbing operator $\mca{V}$ with $r=4$ has $\ga_{\mca{V}}^{(1)} = 2$, so it has scaling dimension $\DD_{\mca{V}} = 4 + \sO(\vareps^2)$.\footnote{Thus the operator $\mca{V}$ satisfies $d-\DD_{\mca{V}} = - \vareps + \sO(\vareps^2)$ so it's irrelevant at the IR fixed point, at least at leading order in the $\vareps$ expansion.}

\subsection{Bounds and sum rules}

We will now use Eq.~\reef{eq:genEV} to prove properties of one-loop anomalous dimensions $\gamma_v^{(1)}$ in general multiscalar CFTs. First, we will argue that the individual $\ga_v^{(1)}$ are  bounded, and second we will study certain averages of anomalous dimensions, namely
\beq
\label{eq:sums}
\expec{\ga^n}^\la_r = \frac{1}{d_{N,r}} \sum_{v=1}^{d_{N,r}} (\ga_v^{(1)})^n
\qquad
\text{for}
\quad
n=1,2.
\eeq
It turns out that sums of the form~\reef{eq:sums} can be expressed in terms of $n$-th order invariants of the coupling $\la_{ijkl}$ and as such they can be related to the bounds derived in Sec.~\ref{sec:boundsMAIN}.

First, let us consider the individual anomalous dimensions $\ga_v^{(1)}$. Setting $B=16\pi^2$, we can simply appeal to~\reef{eq:ab} to find that
\beq
\label{eq:gammabound}
{
|\ga_v^{(1)} | \leq \frac{r(r-1)}{2} \, |\la|.
}
\eeq
Moreover, equality in~\reef{eq:gammabound} only holds for the Wilson-Fisher CFT. To be precise, the bound is saturated for the following potential and operators:
\beq
V(\phi) = \frac{1}{3} (u \cdot \phi)^4
\qaq
\Oo = (u \cdot \phi)^r
\eeq
where $u_i$ can be any unit vector.  In other words, the well-known one-loop anomalous dimensions $\tfrac{1}{6}r(r-1)$ of the Wilson-Fisher fixed point play a special role.

The bound~\reef{eq:gammabound} may be compared to an existing result in the $O(N)$ model. In Ref.~\cite{KEHREIN1993669}, Kehrein et al.\@ showed that \emph{any} operator with $r$ fields (possibly with gradients) in that theory has a bounded anomalous dimension:
\beq
\label{eq:Kehreinbnd}
\hspace{-30mm} \text{$O(N)$ CFT}:
\qquad
0 \leq \ga^{(1)} \leq \frac{r(3r+N-4)}{2(N+8)}\,.
\eeq
If $N$ is large, this result is significantly stronger than~\reef{eq:gammabound}, but formula~\reef{eq:Kehreinbnd} does reflect the $\sim r^2$ scaling at large $r$ of~\reef{eq:gammabound}.

Next, we turn to the averages $\expec{\ga}_r$ and $\expec{\ga^2}_r$ from~\reef{eq:sums}. For specific theories, such sums can be computed explicitly. In appendix~\ref{app:data}, we solve the eigenvalue problem~\reef{eq:genEV} for the case of the Wilson-Fisher model (tensored with $N-1$ free fields) and the $O(N)$ fixed points. For the Wilson-Fisher CFT, we find for instance that
\beq
\label{eq:WFdata}
\expec{\ga}^\text{WF}_r = \frac{r(r-1)}{3N(N+1)}
\qaq
\expec{\ga^2}^\text{WF}_r = \frac{r(r-1)}{9 (N)_{4}} [N^2+6(r-1)^2 +N(6r-7)]
\eeq
where $(x)_n = x(x+1)\dotsm (x+n-1)$ is the Pochhammer symbol. For the $O(N)$ CFT, we have instead
\bsub
\label{eq:ONdata}
\beq
\label{eq:ONlin}
\expec{\ga}^{O(N)}_r = \frac{r(r-1)}{N(N+1)} \frac{N(N+2)}{N+8}
\eeq
and
\begin{multline}
\label{eq:ONsq}
  \expec{\ga^2}^{O(N)}_r = \frac{r(r-1)}{(N)_4} \frac{N(N+2)}{(N+8)^2} \; H(N,r),\\
  H(N,r) = N(N-1) - (13+4N+N^2)r + (11+6N+N^2)r^2.
\end{multline}
\esub
Moreover, the spectrum of the $N$-field hypercubic fixed point was recently analyzed by Antipin and Bersini in Ref.~\cite{Antipin:2019vdg}, where explicit results were given for scalar operators with $r \leq 5$. They find for instance that
\beq
\label{eq:antipinrule}
\expec{\ga}^\text{hypercubic}_r = \frac{2r(r-1)(N-1)}{3N(N+1)}
\quad
\text{for}
\quad
r=2,3,4,5.
\eeq
The quantity $\expec{\ga^2}^\text{hypercubic}_r$ can be computed as well, and the result is printed in Eq.~\reef{eq:hyperdata}.

The above formulas look simple, even though they arise from the complicated-looking eigenvalue problem~\reef{eq:genEV}. Indeed, we claim that both of these quantities can be expressed in terms of the invariants $\ba_k(\la)$ that we have encountered previously:
\beq
\label{eq:sumrules}
\expec{\ga}^\la_r = \frac{r(r-1)}{N(N+1)}\, \ba_0(\la)
\qaq
\expec{\ga^2}^\la_r = \frac{r(r-1)}{(N)_4} \, \mca{S}_{N,r}(\ba_0(\la),|\la|^2)
\eeq
where the function $\mca{S}_{N,r}$ is given by
\beq
\label{eq:quadsumrule}
\mca{S}_{N,r}(x,y) = 4(r-2)(N+r)x + (r-2)(r-3)x^2 - (N+r)(7r-17-N) y
\eeq
using the shorthand notation $x = \ba_0(\la)$, $y = |\la|^2$. The derivation of these identities is straightforward but slightly cumbersome, and we defer the proof of Eq.~\reef{eq:sumrules} to Appendix~\ref{sec:proofsumrules}.  In the derivation of the second sum rule the relation~\reef{eq:quad} was used, so it only holds at fixed points. It's easy to check that~\reef{eq:sumrules} is valid for the Wilson-Fisher, the $O(N)$ and the hypercubic CFTs (for $r \leq 5$). In fact, the first sum rule shows that the formula~\reef{eq:antipinrule} holds for any $r \geq 2$, as was already conjectured in~\cite{Antipin:2019vdg}.

Above, we showed that individual anomalous dimensions $\ga^{(1)}$ were bounded in terms of $|\la|$. In particular, it follows from~\reef{eq:gammabound} and the Rychkov-Stergiou bound~\reef{eq:RSref} that
\beq
\label{eq:roughbound}
\big| \expec{\ga}_r^\la \big| < \frac{ r(r-1)}{4\sqrt{2}} \, \sqrt{N}
\qaq
\big| \expec{\ga^2}_r^\la \big| < \frac{ r^2(r-1)^2}{32} \, N.
\eeq
This is a statement about the spectrum of \emph{any} one-loop CFT in the epsilon expansion, regardless of any information about its coupling $\la_{ijkl}$. However, the bound~\reef{eq:roughbound} is far from optimal, at least in the examples considered above. With some additional work, a sharper statement can be derived.

\begin{figure}[htb]
    \begin{center}
    \hspace{0mm}
    \includegraphics[scale=.6]{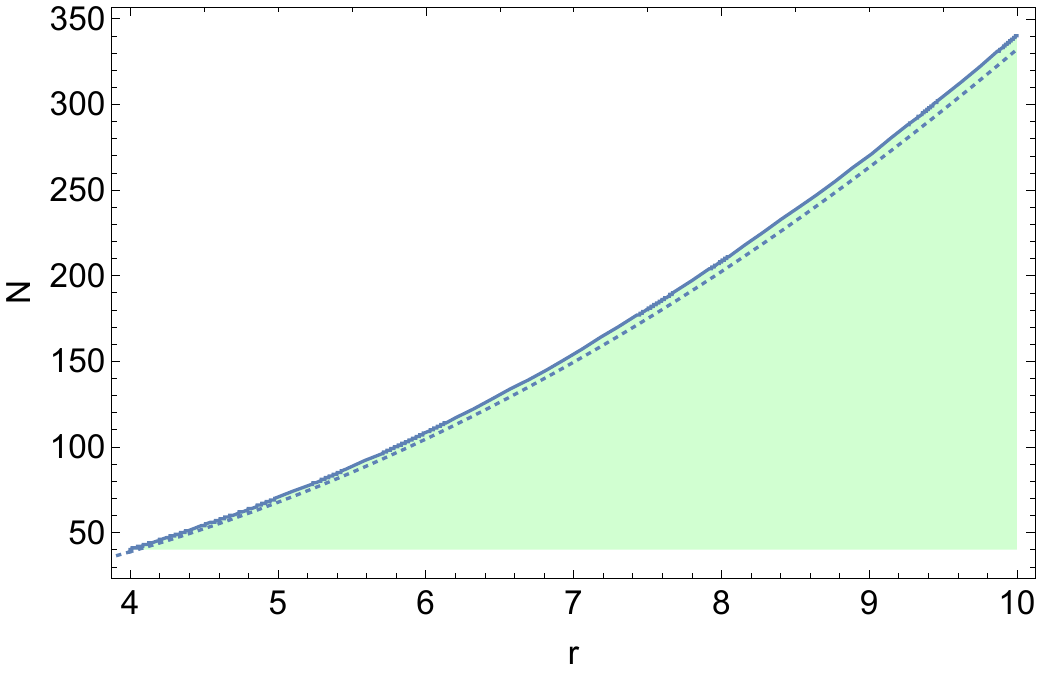}
    \caption{\label{fig:maxregion} The green region shows the values of $r$ and $N$ for which the quantity $\expec{\ga^2}_r$ is bounded from above by its value in the $O(N)$ theory, according to Eq.~\reef{eq:lastleg}. In the white region, it is possible that another CFT has a larger value of $\expec{\ga^2}_r$. The solid blue line follows from the proof in Apppendix~\ref{app:specbnd}; the dotted line is the approximation $N \leq 4 \left(r-9/8\right)^2$ appearing in~\reef{eq:lastleg}.}
  \end{center}
  \end{figure}

\begin{thm}
Consider any one-loop CFT with $N \geq 2$ scalar fields, labeled by a coupling $\la_{ijkl}$. For any $r \in \{2,3,\ldots\}$, the quantity $\expec{\ga}_r^\la$ is bounded from above by its value in the $O(N)$ theory:
\bsub
\label{eq:thmbounds}
\beq
\label{eq:fact1}
\expec{\ga}_r^\la \leq \expec{\ga}_r^{O(N)} \,.
\eeq
For $r=2$, we have in addition
\beq
\label{eq:fact2}
\expec{\ga^2}_2^\la \leq \begin{cases} \expec{\ga^2}_2^{O(N)} &\text{for} \quad N=2,3,4 \\ \frac{1}{4(N+1)} &\text{for}\quad N \geq 5 \end{cases}\;.
\eeq
If $N$ is sufficiently small, the following analog of~\reef{eq:fact1} holds for $r \geq 3$:
\beq
\label{eq:lastleg}
\hspace{-10mm} N \leq 4 \left(r-\frac{9}{8}\right)^2:
\qquad
\expec{\ga^2}_r^\la \leq \expec{\ga^2}_r^{O(N)} < \frac{r(r-1)(r^2 - r + 1)}{(N+5)^2}\,.
\eeq
The domain of $r$ and $N$ for which $\expec{\ga^2}_r$ is maximal at the $O(N)$ fixed point is shown in Fig.~\ref{fig:maxregion}.
\esub
\end{thm}
The above result is a clear improvement compared to the naive bound~\reef{eq:roughbound}: in particular, the average of the anomalous dimensions $\expec{\ga}_r$ scales at most as $1/N$ at large $N$ --- as can be seen from eq.~\reef{eq:ONlin} --- compared to $\sqrt{N}$. We also remark that the restriction on $N$ from~\reef{eq:lastleg} is not very severe; even for $r=3$, it only requires $N \leq 14$ (and in fact the result holds for $N \leq 18$ in that case).

The proof of the bounds~\reef{eq:fact1} is straightforward but long. It relies on the fact that the quantities $\expec{\ga^{1,2}}_r^\la$ can be expressed in terms of the invariants $\ba_0(\la)$ and $|\la|^2$, which we know are restricted to a compact domain $\mca{D}_N \subset \mbb{R}^2$. Therefore
\beq
\label{eq:supformula}
\expec{\ga^2}_r^\la \leq \frac{r(r-1)}{(N)_4} \; \sup_{\mca{D}_N}\, \mca{S}_{N,r}
\eeq
and determining the supremum of $\mca{S}_{N,r}$ is an exercise in calculus. Some details are provided in Appendix~\ref{app:specbnd}.

 Without the restriction on $N$, it is still possible to find bounds on the quantity $\expec{\ga^2}_r^\la$, but such bounds won't involve the $O(N)$ fixed point. For instance, for $r = 3$ and $N \geq 19$ we can show, using the methods of Appendix~\ref{app:specbnd}, that
 \beq
 \label{eq:ga3val}
 \hspace{-20mm}
 N \geq 19:
 \qquad
 \expec{\ga^2}_3^\la \leq \frac{3(N+4)^2}{4(N-4)(N+1)(N+2)}
 \eeq
 which is somewhat larger than $\expec{\ga^2}_3^{O(N)}$ for these values of $N$. The bound~\reef{eq:ga3val} can only be saturated by an isotropic CFT with $\ba_0 = \th N(N+4)/(N-4)$; it is an interesting question whether such a one-loop fixed point exists. Similar bounds can be obtained for $r \geq 4$, but we will not discuss this further.

Using the same ideas, it is possible to derive uniform lower bounds on $\expec{\ga}_r$ and $\expec{\ga^2}_r$. Such lower bounds don't appear to involve any know fixed points, and we will not discuss them in the present work either.

Let us finally comment on the case where $\la_{ijkl}$ is isotropic. In that case, the order-$\vareps$ contribution to the critical exponent $\nu$ is given by the anomalous dimension of the operator $\phi \cdot \phi$, which  can be computed as a special case of~\reef{eq:genEV}. Using this information, the authors of~\cite{Brezin:1973jt} derived a bound on $\nu$ valid for general isotropic theories, comparing it to the value of $\nu$ of the $O(N)$ fixed point.

%% file: sections/complex.tex

So far, we studied in detail the theory of $N$ real scalars $\phi^i$ interacting by means of a  quartic interaction. The landscape of theories with $N$ \emph{complex} scalars is not as well-explored. In this section, we will adapt the methods from Sec.~\ref{sec:boundsMAIN} to analyze two types of Lagrangians with complex degrees of freedom: first, the theory of $N$ complex scalars with a $\phib \phib \phi \phi$ interaction, and second the $U(1)$ gauge theory of $N$ charged scalars, also known as bosonic QED.

\subsection{Interactions with $N$ complex scalars}

To proceed, let us consider the theory of $N$ complex scalars $\phi^i$ and their conjugates $\bar{\phi}^i$ in $4-\vareps$ dimensions, interacting by means of the following quartic interaction:
\beq
\label{eq:compAction}
\msc{L} = |\pd_\mu \phi^i|^2 + \frac{B'}{6} V(\phi^i,\phib^i),
\quad
V(\phi^i,\phib^i) = g_{ijkl} \, \bar{\phi}^i\bar{\phi}^j{\phi}^k{\phi}^l\,.
\eeq
Here $B'$ is an arbitrary constant that will be fixed later. 
It is possible to allow for a more general quartic interaction, including  terms of the form $h_{ijkl} \, \phi^i \phi^j \phi^k \phi^l$, $h'_{ijkl} \, \phib^i \phi^j \phi^k \phi^l$ and their conjugates.  However, the action~\reef{eq:compAction} is invariant under the $U(1)$ symmetry $Q$ that assigns a charge $+1$ to $\phi^i$ and $-1$ to $\bar{\phi}^i$, and we will later gauge precisely this $U(1)$.

By construction, the couplings $g_{ijkl}$ are invariant under the $\mbb{Z}_2 \times \mbb{Z}_2$ symmetry
\beq
\label{eq:compsymm}
g_{ijkl} = g_{jikl} = g_{ijlk}
\eeq
contrary to the full $\mca{S}_4$ permutation invariance of the real coupling $\la_{ijkl}$. In addition, the reality of $\msc{L}$ enforces that
\beq
\label{eq:reality}
g_{ijkl}^* = g_{klij}.
\eeq
It's easy to show that tensors satisfying~\reef{eq:compsymm} and~\reef{eq:reality} are parametrized by $\tfrac{1}{4}N^2(N+1)^2$ real coefficients.

Setting $B' = 24\pi^2$, the one-loop beta function of $g_{ijkl}$ is given by
\beq
\label{eq:complexbeta}
\beta(g)_{ijkl}\, \bar{\phi}^i\bar{\phi}^j{\phi}^k{\phi}^l = (-g_{ijkl} + g_{ijmn} g_{mnkl} + 4 g_{imkn} g_{jnlm}) \bar{\phi}^i\bar{\phi}^j{\phi}^k{\phi}^l.
\eeq
Even though $V$ is not manifestly positive for an arbitrary coupling $g_{ijkl}$, it is indeed positive at fixed points. To check this, we simply rewrite the above equation as
\beq
\beta(g) = 0
\quad
\Rightarrow
\quad
V(\phi,\phib) = \text{tr}(L^\dagger L) + 4\, \text{tr}(M^\dagger M) \geq 0
\eeq
with
\beq
L_{ij}(\phi) = g_{ijmn} \, \phi^m \phi^n
\qaq
M_{ij}(\phi,\phib) = g_{i m jn} \, \phib^m \phi^n.
\eeq

As an example, take the most symmetric interaction of the form~\reef{eq:compAction}:
\beq
\label{eq:UNfp}
V(\phi^i,\bar{\phi}^j) = \frac{1}{N+4}\, (\phi^i \bar{\phi}^i)^2\,.
\eeq
This potential is invariant under a $U(N)$ symmetry that acts on the fields as follows:
\beq
\label{eq:UNaction}
\hspace{-25mm}
U(N) \ni \mca{R}:
\qquad
\phi^i \mapsto \mca{R}\ud{i}{j} \phi^j,
\quad
\bar{\phi}^i \mapsto (\mca{R}^*)\ud{i}{j} \bar{\phi}^j = (\mca{R}^\dagger)\du{j}{i} \bar{\phi}^j.
\eeq
The prefactor $1/(N+4)$ in~\reef{eq:compAction} is fixed by imposing that~\reef{eq:UNfp} is a fixed point. For $N=1$, this is the theory of a single doublet with a $(\phi \phib)^2$ interaction (sometimes known as the XY model). As a side note, the $U(N)$ fixed point~\reef{eq:UNfp} is equal to the $O(2N)$ fixed point from~\reef{eq:ONfp}, as can be seen by decomposing $\phi^i$ and $\bar{\phi}^i$ in terms of 2$N$ real degrees of freedom.

\subsection{Invariants and bounds}
\label{bound-complex}

Just as before, we can try to restrict the space of zeroes of the one-loop beta function~\reef{eq:complexbeta}. By analogy with the result~\reef{eq:WFb}, we can manipulate Eq.~\reef{eq:complexbeta} to show that at a fixed point
\beq
\label{eq:WFbc}
g_{ijkl} = 0
\qor
\norm{g} \geq \frac{1}{5}.
\eeq
The proof of this inequality is given in Appendix~\ref{sec:complexvee}. This bound is optimal, since the $U(1)$ fixed point from Eq.~\reef{eq:UNfp} has $\norm{g} = \tfrac{1}{5}$. 

To obtain more refined bounds, we can introduce three projection operators $\mca{Q}_i$ as follows:
\vspace{-6mm}
\bsub
\label{eq:sunproj}
\begin{align}
  \mca{Q}_{1}(g)_{ijkl} &= \frac{1}{N(N+1)}\, (\dd_{ik} \dd_{jl} + \dd_{il} \dd_{jk}) g_{mnmn} \\
  \mca{Q}_{2}(g)_{ijkl} &= \frac{1}{N+2} \left[\dd_{i k} g_{jm lm}+ \text{3 terms} - (2N+2)\mca{Q}_{1} (g)_{ijkl}\right] \\
  \mca{Q}_{3}(g) &= g - \mca{Q}_{1}(g) - \mca{Q}_{2}(g)
\end{align}
\esub
which indeed obey $\mca{Q}_i \mca{Q}_j = \dd_{ij} \mca{Q}_i$. Associated to these projectors, we can define invariants
\beq
\bb_1(g) = g_{ijij},
\quad
\bb_\ell(g) = g_{klij} \mca{Q}_\ell(g)_{ijkl} = \norm{\mca{Q}_i(g)}^2
\quad
\text{for}
\;
\ell=1,2.
\eeq
Owing to the reality condition~\reef{eq:reality} $\bb_1(g) \in \mbb{R}$, and by construction $\bb_2(g), \bb_3(g) \geq 0$. These invariants are related to the norm $\norm{g}^2 = g_{ijkl}g_{klij}$ as follows:
\beq
\label{eq:invrelc}
\norm{g}^2 = \frac{2}{N(N+1)}\, \bb_1(g)^2 + \bb_2(g) + \bb_3(g).
\eeq
For example, the $U(N)$ fixed point~\reef{eq:UNfp} has invariants
\beq
\label{eq:UNinv}
g = g^{U(N)}:
\qquad
\bb_1(g) = \frac{N(N+1)}{2(N+4)},
\quad
\bb_2(g) = 0
\qaq
\norm{g}^2 = \frac{N(N+1)}{2(N+4)^2}.
\eeq

In passing, remark the complex analog of the isotropy requirement~\reef{eq:tracecond} is
\beq
\label{eq:compIso}
g_{ikjk} = w \cdot \dd_{ij},
\quad
w \in \mbb{R}.
\eeq
If Eq.~\reef{eq:compIso} is satisfied,  $\bb_2(g) = 0$. In particular, the $U(N)$ fixed point has this isotropy property. Just like in the real case, we will not impose~\reef{eq:compIso}.

Let us make a brief comment on the group-theoretical meaning of the projectors $\mca{Q}_\ell$. The group $U(N)$ acts on the tensor $g_{ijkl}$ through
\beq
U(N) \ni \mca{R}:
\quad
g_{ijkl} \mapsto g^\mca{R}_{ijkl} = (\mca{R}^\dagger)\du{i}{i'} (\mca{R}^\dagger)\du{j}{j'} \mca{R}\ud{k'}{k} \mca{R}\ud{l'}{l} \; g_{i'j'k'l'}
\eeq
cf.\@ the action~\reef{eq:UNaction} on the fields $\phi^i$, $\phib^i$. Under this action the tensor $g_{ijkl}$ is reducible: its irreducible components can be identified with a trivial representation, the adjoint rep of dimension $N^2-1$ and a third representation of dimension $\tfrac{1}{4}N^2(N-1)(N+3)$, all of which are self-conjugate. These three irreps are precisely the image of the projectors $\mca{Q}_1$, $\mca{Q}_2$ and $\mca{Q}_3$. 

In order to find an additional relation between the invariants that holds at fixed points, let us introduce the matrix $\ga_{ij} = g_{ikjk}$, which is Hermitian ($\ga^\dagger = \ga$). From~\reef{eq:complexbeta}, it follows that
\beq
\label{eq:miley}
\bb_1(\beta(g))= - \bb_1(g) + 3\norm{g}^2 + 2 \text{tr}(\ga^2).
\eeq
At the same time,
\beq
\bb_2(g) = \frac{4}{N+2}\! \left( \text{tr}(\ga^2) - \frac{1}{N} \, \bb_1(g)^2 \right)\!.
\eeq
Consequently, at a fixed point we can write
\beq
\label{eq:quadcomplex}
\beta(g) = 0
\quad
\Rightarrow
\quad
\frac{2}{3N}\, \bb_1(g)\!\left(\frac{N}{2} - \bb_1(g)\right) = \norm{g}^2 + \frac{1}{6}(N+2) \bb_2(g).
\eeq

We are now in a very similar situation as before: the (in)equalities~\reef{eq:WFbc}, ~\reef{eq:invrelc} and~\reef{eq:quadcomplex} are nearly identical to their real counterparts~\reef{eq:WFb}, ~\reef{eq:invrel} and~\reef{eq:quad}, the only difference being that some coefficients differ. It is therefore straightforward to show that e.g.\@ $\bb_1(g)$ and $\norm{g}^2$ can be restricted to a compact domain. We will not provide a complete derivation, since it would be almost identical to Sec.~\ref{sec:realinv}. Instead, we will provide a summary of the results.

\begin{thm}
Given the action~\reef{eq:compAction} with $N \geq 1$ fields, if $g_{ijkl}$ describes a non-trivial CFT at one loop, then
\bsub
\beq
\label{eq:b1bnd}
\sigma_N^{-} < \bb_1(g) \leq \sigma_N^+
\eeq
with
\beq
\sigma_N^+ = \frac{N(N+1)}{2(N+4)}
\qaq
\sigma_N^- = \frac{N}{4} \left[ 1-\sqrt{1-\frac{24}{25N}}\right].
\eeq
\esub
The upper bound in~\reef{eq:b1bnd} is optimal, because $\sigma_N^+$ coincides with $\bb_1$ of the $U(N)$ CFT. Inside this interval,
\beq
\label{eq:gsqba}
\max \left\{ \frac{1}{25}, \, \frac{2 \bb_1(g)}{N+8}  \left(1-\frac{\bb_1(g)}{N+1}\right) \right\} \leq \norm{g}^2 \leq \frac{2}{3N} \, \bb_1(g) \!\left( \frac{N}{2} - \bb_1(g)\right).
\eeq
In particular
\beq
\label{eq:complexRS}
\norm{g}^2 \leq \begin{cases} \norm{g^{U(N)}}^2 = \tfrac{1}{25} &\text{for } \; N=1 \\ \tfrac{1}{24}N &N \geq 2
\end{cases}\;.
\eeq
\end{thm}
\noindent The inequality~\reef{eq:complexRS} is the counterpart of the Rychkov-Stergiou bound~\reef{eq:RSbound} and its refinement~\reef{eq:RSref} in the case of $N$ real scalars. 

Let us make a final comment on the role of charge conjugation $\mca{C}$, the $\mbb{Z}_2$ symmetry that exchanges fields and their conjugates, $\mca{C} :  \phi^i \lra \bar{\phi}^i$. From the point of view of~\reef{eq:compAction}, $\mca{C}$ is a discrete symmetry that may be imposed, although we do not do so in this work. Concretely, charge conjugation implies that $g_{ijkl} = g_{klij}$ or equivalently $g_{ijkl} \in \mbb{R}$, recalling the reality condition~\reef{eq:reality}. From a practical point of view, imposing $\mca{C}$ reduces the number of real couplings to $\tfrac{1}{8}N(N+1)(N^2 + N + 2)$. It is an open question whether $\mca{C}$ puts interesting constraints on the space of theories --- we only remark that ``simple'' CFTs like the $U(N)$ fixed point~\reef{eq:UNfp} are manifestly invariant under $\mca{C}$. 

\subsubsection{Comparison to real case}

As previously mentioned, the theory with $N$ complex scalars~\reef{eq:compAction} obeying the reality condition~\reef{eq:compsymm} is a special case of the Lagrangian~\reef{eq:genLag} with $2N$ real scalar fields. Explicitly, one can write the complex fields $\phi^i$ in terms of real fields $\psi^{i,\a}$ with $\a = 1,2$ as
\beq
\begin{pmatrix} \phi^i \\ \phib^i \end{pmatrix}  = \frac{\psi^{i,1} \pm i \psi^{i,2}}{\sqrt{2}}
\eeq
such that the $\psi^{i,\a}$ are canonically normalized. The $U(1)$ global symmetry of the action~\reef{eq:compAction} translates to an $SO(2)$ global symmetry acting on the doublets $\psi^{i,\a}$, so the theory in question is not the most general multiscalar theory.\footnote{In fact, the complex theory is specified by $\sim \tfrac{1}{4}N^4$ couplings, whereas a theory with $2N$ real fields is described by $\sim \tfrac{2}{3}N^4$ couplings.}

In particular, any statement about fixed points with complex scalars should be a refinement of theorems described previously for real fixed points. Given the normalizations $B = 16\pi^2$ and $B' = 24\pi^2$, the invariants $\norm{g}^2$ and $\bb_j(g)$ are related to $\norm{\la}^2$ and $\ba_j(\la)$ via
\beq
\norm{g}^2 = \tfrac{1}{6} \norm{\la}^2,
\quad
\bb_1(g) = \tfrac{1}{4} \ba_0(\la),
\quad
\bb_2(g) = \tfrac{1}{6} \ba_2(\la)
\qaq
\bb_3(g) = \tfrac{1}{6} \ba_4(\la).
\eeq
It follows that the upper bound on $\norm{g}^2$ from~\reef{eq:complexRS} is identical to~\reef{eq:RSref}, but the lower bound $\norm{g} \geq \tfrac{1}{5}$ translates to $\norm{\la} \geq \tfrac{\sqrt6}{5} \approx 0.49$, which is stronger than $\norm{\la} \geq \tfrac{1}{3}$. This is not surprising, as these two bounds correspond to different CFTs, the $O(2)$ versus the Ising model. The remaining constraints on $\norm{g^2}$ from~\reef{eq:gsqba} are identical to those obtained in the real case. Finally, the upper bound $\sigma_N^{+}$ on $\bb_1(g)$ from~\reef{eq:b1bnd} agrees with the upper bound $\varrho_{2N}^+$ on $\ba_0(\la)$ from~\reef{eq:a0bd}, although the lower bound $\sigma_N^{-}$ is sharper than its counterpart $\varrho_{2N}^{-}$.

\subsection{Bosonic QED}

In this subsection we discuss what happens to the Lagrangian~\eqref{eq:compAction} if we gauge the $U(1)$ symmetry under which $\phi^i$ resp.\@ $\phib^i$ have charges $\pm 1$. The resulting action describes bosonic versions of QED in $4-\vareps$ dimensions, allowing for the most general coupling $g_{ijkl}$ that is consistent with gauge invariance and unitarity.  Concretely, we minimally couple the matter fields $(\phi^i,\phib^i)$ from~\reef{eq:compAction} to a photon $A_\mu$:
\beq
\label{eq:compActionG}
\msc{L} = |D_\mu \phi^i|^2 + 4\pi^2 \, V(\phi,\phib) + \frac{1}{4} F_{\mu \nu} F^{\mu \nu},
\quad
V(\phi,\phib) =  g_{ijkl} \, \phib^i\phib^j\phi^k\phi^l
\eeq
where $D_{\mu } = \partial_{\mu} + i e A_{\mu}$ is the gauge covariant derivative and $F_{\mu \nu} = \pd_\mu A_\nu - \pd_\nu A_\mu$. For later convenience, we can redefine the gauge coupling as follows:
\beq
\a \ldef \frac{e^2}{24\pi^2}.
\eeq
The one-loop beta functions of $\a$ and $g_{ijkl}$ are given by\footnote{These beta functions can be derived starting from those of scalar QED in four dimensions.}
\bsub
\begin{align}
\beta(\a) &= - \vareps \a + N \a^2\\
  {\beta}(g)_{ijkl}  \phib^i\phib^j\phi^k\phi^l  &= \left[ -(\vareps +18\a) g_{ijkl} + (g \flat g)_{ijkl} + 4 (g \sharp g)_{ijkl} \right]\! \phib^i\phib^j\phi^k\phi^l + 54 \a^2 (\phi \cdot \phib)^2
\end{align}
\esub
where $\flat$ and $\sharp$ are two tensor contractions defined in Eq.~\reef{eq:music}. The gauge coupling $\a$ has two possible values at a fixed point: $\a = 0$ or $\a = \a_\star \ldef \vareps/N$. If $\a = 0$, the photon decouples and the scalars interact according to the ungauged interaction~\reef{eq:compAction}. If $\a = \a_\star$ however, the beta function of the scalar couplings is modified. Rescaling $g_{ijkl}$ by a factor $1/\vareps$, these modified beta functions are given by
\beq
\label{eq:starbeta}
\beta(g)_{ijkl} = \beta_0(g)_{ijkl} - \frac{18}{N}\,  g_{ijkl} + \frac{54}{N^2}\,  T_{ijkl}\,,
\quad
T_{ijkl} \ldef \half(\dd_{ik} \dd_{jl} + \dd_{il} \dd_{jk})
\eeq
where $\beta_0(g)$ denotes the ungauged one-loop beta function~\reef{eq:complexbeta}. Remark that in the limit $N \to \infty$, the two terms of order $1/N$ and $1/N^2$ vanish, and the ungauged beta function $\beta_0(g)$ is recovered.

As an example, we can look for maximally symmetric fixed points, having $PSU(N)$\footnote{The global symmetry group is not equal to $U(N)$, since the $U(1)$ factor is gauged and must be modded out. In addition, Eq.~\reef{eq:symmsols} has a $\mbb{Z}_2$ charge conjugation symmetry $\mca{C}$.} as a global symmetry group:
\bsub
\label{eq:symmsols}
\beq
V(\phi,\phib) = c \, (\phi^i \phib^i)^2,
\quad
c \in \mbb{R}.
\eeq
If the number of flavors $N$ is larger than $183$, there are two unitary fixed points with different values of $c$, namely\footnote{These solutions have the following invariants:
\[
\bb_1(g^\pm) = \frac{c_\pm}{2} N(N+1),
\quad
\bb_2(g^\pm) = \bb_3(g^\pm) = 0
\qaq
\norm{g^\pm}^2 = \frac{c_\pm^2}{2} \, N(N+1).
\]
}
\beq
c_\pm = \frac{1}{2N(N+4)} \left( N +18 \pm \sqrt{N^2 - 180N -540} \right)\!.
\eeq
\esub
The two solutions are referred to as bQED${}_+$ and bQED in~\cite{Benvenuti:2019ujm}: bQED${}_+$ is known as the abelian Higgs or the non-compact CP$^{N-1}$ model, whereas bQED is unstable. The fact that the solutions~\reef{eq:symmsols} only exist for $N \geq 183$ (at leading order in $\vareps$!) has been known for a long time~\cite{Halperin:1973jh}. For even values of $N$, Refs.~\cite{Benvenuti:2018cwd,Benvenuti:2019ujm} also describe two families of one-loop CFTs with symmetry group $SU(\th N) \times SU(\th N)$, which exist for $N \geq 197$.

These results suggest that it is difficult to construct one-loop fixed points of the Langrangian~\reef{eq:compActionG} with a large symmetry group but a finite number of flavors, say $N \lesssim 100$.  However, it could be possible that there exist CFTs at small $N$ that have a small or even trivial symmetry group. In what follows we will show that this is not the case, by proving that there are no one-loop CFTs at all with $N < 183$ flavors. To do so, we employ the by now familiar strategy of carefully analyzing relations between the invariants $\bb_\ell(g)$ and $\norm{g}^2$. For example, we can apply the invariant $\bb_1$ to~\reef{eq:starbeta}, using the fact that $\bb_1(T) = \th N(N+1)$ as well as~\reef{eq:miley}. This leads to the identity
\beq
\label{eq:bstst}
\beta_\star(g) = 0
\quad
\Rightarrow
\quad
\frac{2}{3N} \, \bb_1(g) \left(\frac{N+18}{2} - \bb_1(g) \right) = \norm{g}^2 + \frac{1}{6}(N+2)  \bb_2(g) + \frac{9(N+1)}{N}.
\eeq
Using~\reef{eq:invrelc}, $\norm{g}^2$ can be eliminated in favor of $\bb_3(g)$:
\beq
\frac{\bb_1(g)}{3\sigma_N^+} \left(\frac{N+18}{N} \, \sigma_N^+ - \bb_1(g)\right) - \frac{1}{6}(N+8)\bb_2(g) -\bb_3(g) = \frac{9(N+1)}{N}.
\eeq
By construction, $\bb_2(g)$ and $\bb_3(g)$ are positive. Therefore we must have
\beq
\frac{9(N+1)}{N} \leq \; \frac{1}{3\sigma_N^{+}} \; \sup_{x \in \mbb{R}} \; x \left(\frac{N+18}{N} \, \sigma_N^+ - x\right) = \frac{(N+1)(N+18)^2}{24N(N+4)}.
\eeq
But this inequality can only hold for
\beq
\label{eq:Nbound}
N \geq  90 + 24\sqrt{15} \approx 182.95
\eeq
as we claimed above.
  
\begin{figure}[!htb]
    \begin{center}
    \hspace{0mm}
    \includegraphics[scale=.8]{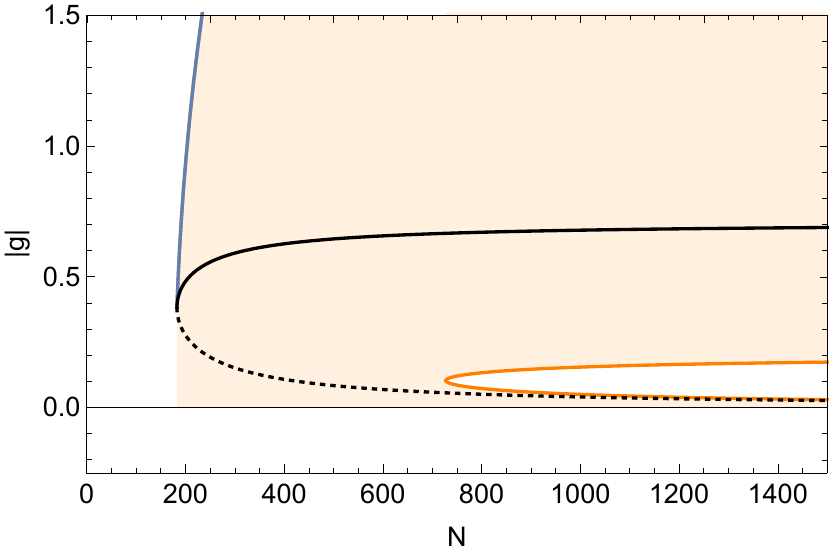}
    \caption{\label{fig:complexPlot} Orange region: the values that $\norm{g}$ can take at fixed points, as a function of the number of fields $N$. The blue curve is the upper bound from~\reef{eq:bluebound}; the orange curves are the two branches of Eq.~\reef{eq:orangebound}. The solid and dotted black curves are the two CFTs from Eq.~\reef{eq:symmsols}.}
  \end{center}
   \end{figure}

More quantitatively, we can obtain upper and lower bounds on $\norm{g}$. To wit, if we use that $\bb_2 \geq 0$ then it follows from~\reef{eq:bstst} that
\begin{align}
  \norm{g}^2 &\leq \frac{2}{3N} \, \bb_1(g) \left(\frac{N+18}{2} - \bb_1(g) \right) - \frac{9(N+1)}{N} \nn
  \\ &\leq - \frac{9(N+1)}{N} +  \frac{2}{3N}\, \sup_{x \in \mbb{R}}  \, x \left(\frac{N+18}{2} - x \right) \nn \\
  &= \frac{N}{24} \left(1-\frac{180}{N}+\frac{108}{N^2}\right). \label{eq:bluebound}
\end{align}
Notice that in the limit $N \to \infty$, the inequality~\reef{eq:bluebound} is asymptotically equal to~\reef{eq:complexRS} from the ungauged case. Finally, we can analyze the beta function~\reef{eq:starbeta} directly, rewriting it as
\beq
\left(\frac{N+18}{N}\right)g = \mrm{P}(g \flat g + 4 g\sharp g)  + \frac{54}{N^2} \, T,
\quad
\mrm{P}(A)_{ijkl} \ldef \tfrac{1}{4}(A_{ijkl} + A_{jikl} + A_{ijlk} + A_{jilk}).
\eeq
Here $\mrm{P}$ is nothing but a projection operator that enforces~\reef{eq:compsymm}.
Using the triangle inequality, the inequalities from Sec.~\ref{sec:complexvee} and the fact that $\norm{T}^2 = \th N(N+1)$,  it follows that
\beq
\label{eq:orangebound}
\left(\frac{N+18}{N}\right) \norm{g} \leq 5 \norm{g}^2 + 27 \sqrt{\frac{2(N+1)}{N^3}}.
\eeq
For $N \geq 728$ this leads to a new constraint on $\norm{g}$, namely that either $ \norm{g} \leq \xi_{-}(N)$ or $\norm{g} \geq \xi_{+}(N)$ for two algebraic functions $\xi_\pm(N)$ that can be determined from~\reef{eq:orangebound}. Approximating $\xi_\pm(N)$, it can be shown that
\beq
\hspace{-20mm}
N \geq 728:
\qquad
0 \leq \norm{g} < \frac{74}{N}
\qor
\frac{1}{5} - \frac{70}{N} < \norm{g} < \sqrt{\frac{1}{24}N}.
\eeq
In Fig.~\ref{fig:complexPlot}, the possible values of $\norm{g}$ consistent with Eqs.~\reef{eq:bluebound} and~\reef{eq:orangebound} are shown as a function of $N$.

%% file: sections/discussion.tex

In this work we showed that one-loop fixed points with $N$ real or complex scalars are strongly constrained by unitarity alone. Hopefully, these results provide a stepping stone towards further numerical work along the lines of~\cite{Codello:2020lta}. In our work we used linear and quartic invariants of the coupling $\la_{ijkl}$, but it is also possible to construct higher-order $O(N)$ invariants. If it's possible to improve our bounds using e.g.\@ cubic or quartic invariants, it would certainly be worthwhile to do so. In addition, it would be interesting to find a deeper explanation why there seem to be many fixed points close to the $O(4)$ theory, as we mentioned in Sec.~\ref{sec:realinv}. 

We expect that the methods used in this work generalize to many other situations. For instance, it is possible to study  $\phi^n$-type interactions with $n \neq 4$, which are marginal in $d = 2n/(n-2)$ --- see e.g.~\cite{Osborn:2017ucf} for a discussion of fixed points with $n=3,6$ or~\cite{Codello:2018nbe} for generic even $n$ and $n=3$. Likewise, one can add $N'$ fermions and attempt to constrain fixed points with quartic and Yukawa couplings. In the latter case, it could be interesting to hunt for supersymmetric fixed points.

In the second part of our work, we studied averages of anomalous dimensions of operators built out of $r$ fields without gradients. For one, it is an obvious question to extend these results to spinning composite operators. Moreover, the quantities $\expec{\ga^{1,2}}$ are similar to averages of anomalous dimensions and OPE coefficients that are encountered in analytic bootstrap computations (see e.g.~\cite{Carmi:2020ekr}), and it would be interesting to make contact with these results, either in general or for specific theories.

\subsubsection*{Acknowledgements}

This research was supported in part by Perimeter Institute for Theoretical Physics. Research at Perimeter Institute is supported by the Government of Canada through the Department of Innovation, Science and Economic Development Canada and by the Province of Ontario through the Ministry of Research, Innovation and Science. CT acknowledges support from the Agence Nationale de la Recherche (ANR) grant Black-dS-String (ANR-16-CE31-0004) and the Physics/f NWO grant 680-91-005.
MH is supported by the Simons Foundation grant 488649 (Simons Collaboration on the Nonperturbative Bootstrap) and by the Swiss National Science Foundation through the project 200021-169132 and through the National Centre of Competence in Research SwissMAP. 
The authors thank Chris Beem, Yannick Bonthonneau, Zachary Fisher, Davide Gaiotto, Gianluca Inverso, Theo Johnson-Freyd, Carlo Meneghelli, Slava Rychkov and Andy Stergiou for discussions and correspondence. In particular we thank A.\@ Stergiou for sharing a draft of~\cite{Osborn:2020vya} before publication.

%% file: sections/appendix.tex

\section{Proof of equations~\reef{eq:ab} and~\reef{eq:saturated}}
\label{AppB}

In this appendix, we prove the two equations~Eqs.~\reef{eq:ab},~\reef{eq:saturated} from the main text. To prove the first inequality, we remark that
\beq
(A \vee B)_{i_1 \dotsm i_r} = C_{(i_1 \dotsm i_r)},
\quad
C_{i_1 \dotsm i_r} = A_{kl i_1 i_2} B_{i_3 \dotsm i_r kl}.
\eeq
The tensor $C$ is simply a contraction of $A$ and $B$ which equals $A \vee B$ after symmetrizing its indices. Therefore
\bsub
\begin{align}
  \norm{A \vee B}^2 &\leq \norm{C}^2 \\
  &= \sum_{\{i_\a\}} \big(\sum_{kl} A_{kl i_1 i_2} B_{i_3 \dotsm i_r kl}\big)^2 \label{eq:cs1} \\
  &\leq  \sum_{\{i_\a\}} \big(\sum_{kl} A_{kl i_1 i_2}\big)^2 \sum_{mn} \big(B_{i_3 \dotsm i_r mn}\big)^2 \label{eq:cs2}\\
  &=\norm{A}^2 \norm{B}^2
\end{align}
\esub
as announced. In passing from~\reef{eq:cs1} to~\reef{eq:cs2}, we used the Cauchy-Schwarz inequality. For that inequality to be saturated, necessarily for all $i_1,\ldots,i_r$ there exists a constant $\mu(i_1,\ldots,i_r)$ such that
    \beq
    A_{i_1 i_2 mn} = \mu(i_1,\ldots,i_r) B_{i_3 \dotsm i_r mn}
    \eeq
    for all $m,n$. To interpret this equation, we can think of $A$ resp.\@ $B$ as the coefficients of a linear map $\msc{A} : \mbb{R}^2 \to (\mbb{R}^2)^*$ resp.\@ $\msc{B} : \mbb{R}^2 \to (\mbb{R}^{r-2})^*$. These maps can be represented by matrices. The previous equation simply implies that all of the rows of $\msc{A}$ and $\msc{B}$ are proportional to one another. Hence there exist symmetric matrices $A', M$ and a symmetric rank-$(r-2)$ tensor $B'$ such that
    \beq
    A = A' \otimes M
    \qaq
    B = B' \otimes M.
    \eeq
    We will now prove two facts:
    \begin{itemize}
    \item[(i)] any symmetric rank-4 tensor $A$ of the above form factorizes as $A = c\, u^{\otimes 4}$ for some constant $c$, and 
    \item[(ii)] if a symmetric rank-$k$ tensor $T$ can be written as $T = T' \otimes u$ for some vector $u_i \neq 0$, then $T = c' \, u^{\otimes k}$ for some constant $c' \in \mbb{R}$.
    \end{itemize}
    Setting $B = T$ and $B' = T' \otimes u$, these two facts yield the desired result.
    
    In order to establish (i), notice that the symmetry of $A$ in its indices requires that $A'$ is proportional to $M$. Hence $A_{ijkl} = c  \, M_{ij} M_{kl}$ for some constant $c$. But the fact that $A$ is symmetric under all permutations also implies that $A_{ijkl} = A_{lijk}$ i.e.\@ $M_{ij} M_{kl} = M_{il} M_{kl}$. By taking traces, this implies that $M^2 - \text{tr}(M)M = 0$. Yet $M$ is diagonalizable (since it's symmetric), and the last equation shows that any non-zero eigenvalue $\nu$ must satisfy $\nu = \text{tr}(M)$. This can only be satisfied if $M$ has at most one non-zero eigenvalue, or in other words the image of $M$ is at most one-dimensional. This establishes that there exists a vector $u_i$ such that $M_{ij} = \pm u_i u_j$.

    The proof of (ii) proceeds by induction. For $k=1$ it's a tautology. Suppose that the claim holds for all $k \leq n$. For $k=n+1$, the symmetry of $T$ implies that
    \beq
    T'_{i_1 \dotsm i_{n}} u_{i_{n+1}} =  T'_{i_1 \dotsm i_{n-1} i_{n+1}} u_{i_n}
    \eeq
    for all indices $i_1,\ldots,i_{n+1}$. By contracting with $u^{i_{n+1}}$, it follows that
    \beq
    T'  = T'' \otimes u
    \quad
    \text{for}
    \quad
    T''_{i_1 \dotsm i_{n-1}} \ldef \frac{1}{|u|^2} \, T'_{i_1 \dotsm i_{n-1} m} u^m.
    \eeq
    But $T'$ has rank $k=n$, so it $T'$ factorizes completely. This completes the proof.

  \section{Sum rules for the Wilson-Fisher, $O(N)$ and hypercubic CFTs}
  \label{app:data}

In what follows we will compute the quantities $\expec{\ga}_r$ and $\expec{\ga^2}_r$ for some infinite families of known fixed points. To this end, it will be useful to write an index-free version of~\reef{eq:genEV}. To wit, if we let $\Oo = v_{i_1 \dotsm i_r} \phi^{i_1} \dotsm \phi^{i_r}$ and $V(\phi) = \la_{ijkl} \phi^i \phi^j \phi^k \phi^l$, then~\reef{eq:genEV} becomes
\beq
\frac{1}{24} V_{,ij}(\phi) \, \Oo_{,ij}(\phi) = \gamma_\Oo^{(1)} \, \Oo(\phi)
\eeq
writing $f_{,ij}(\phi) = \pd_i \pd_j f(\phi)$.
To be clear, here the $\phi^i$ are treated simply as auxiliary vectors, and not as operators, so there is no need to normal order.

For example, we can consider the Wilson-Fisher theory tensored with $N-1$ free fields $V(\phi) = \tfrac{1}{3} \phi_1^4$, which satisfies
\beq
V_{,ij}(\phi) = 4 \dd_{i,1} \dd_{j,1} \phi_1^2.
\eeq
Any polynomial of degree $r$ can be written as
\beq
k = 0,\ldots,r:
\quad
\Oo^{(k)}(\phi) = \phi_1^k \, P_{r-k}(\phi_2,\ldots,\phi_N),
\quad
\text{multiplicity} = d_{N-1,r-k}.
\eeq
Such a polynomial obeys
\beq
\frac{1}{24}V_{,ij}(\phi)  \Oo_{,ij}^{(k)}(\phi) = \frac{1}{6} \phi_1^2 \cdot k(k-1) \phi_1^{k-2} P_{r-k}(\phi_2,\ldots,\phi_N) = \frac{k(k-1)}{6} \, \Oo^{(k)}(\phi).
\eeq
Consequently, the sums over anomalous dimensions are given by
\bsub
\beq
\expec{\ga}^{\text{WF}}_r = \frac{1}{d_{N,r}} \sum_{k=0}^n d_{N-1,r-k} \, \frac{k(k-1)}{6} = \frac{r(r-1)}{3N(N+1)}
\eeq
and
\beq
\expec{\ga^2}^{\text{WF}}_r =  \frac{1}{d_{N,r}} \sum_{k=0}^n d_{N-1,r-k} \, \left(\frac{k(k-1)}{6}\right)^2  = \frac{r(r-1)}{9(N)_4} \left[N^2 + 6(r-1)^2 + N(6r-7) \right]
\eeq
\esub
as shown in Eq.~\reef{eq:WFdata}.

For the $O(N)$ model, any rank-$r$ polynomial can be written in the form
\beq
k=0,\ldots,\floor{r/2}:
\quad
\Oo^{(k)}(\phi) = (\phi \cdot \phi)^k \, P_{r-2k}(\phi)
\eeq
where $P_\ell(\phi)$ is a harmonic polynomial of degree $\ell = r-2k$.  The number of such polynomials is
\beq
\label{eq:TSTmult}
d'_{N,\ell} = d_{N,\ell} - d_{N,\ell-2}.
\eeq
This decomposition reflects the decomposition of symmetric tensors in terms of {traceless} symmetric tensors. Now
\beq
\frac{1}{24} V_{,ij}(\phi) = \frac{1}{2(N+8)} ( \dd_{ij} \phi \cdot \phi + 2 \phi_i \phi_j )
\eeq
and
\begin{multline}
\Oo^{(k)}_{,ij}(\phi) = \big[ 4k(k-1) (\phi \cdot \phi)^{k-2}  \phi_i \phi_j  +  2k \dd_{ij} (\phi \cdot \phi)^{k-1} \\+
2k (\phi \cdot \phi)^{k-1} (\phi_i \pd_j + \phi_j \pd_i) + (\phi \cdot \phi)^k \pd_i \pd_j \big] P_{r-2k}(\phi)
\end{multline}
from which it follows that
\beq
\frac{1}{24} V_{,ij}(\phi) \Oo^{(k)}_{,ij}(\phi) = \ga_{r,k}^{(1)} \, \Oo^{(k)}(\phi),
\quad
\ga^{(1)}_{r,k} = \frac{r(r-1)+k(N+2r-2)-2k^2}{N+8}.
\eeq
Here we have used that $\pd^2 P_{\ell} = 0$ and $(\phi \cdot \pd) P_\ell = \ell P_\ell$. Summing over these results with the correct multiplicities from Eq.~\reef{eq:TSTmult}, we have
\beq
\expec{\ga^n}^{O(N)}_r = \frac{1}{d_{N,r}} \sum_{k=0}^{\floor{r/2}} d'_{N,r-2k} \left(\ga_{r,k}^{(1)}\right)^n
\eeq
which leads to Eq.~\reef{eq:ONdata} in the main text. 

Finally, for the hypercubic fixed point with symmetry group $\mbb{Z}_2^N \rtimes \mca{S}_N$, anomalous dimensions for all operators with $r=2,3,4,5$ have been computed in Ref.~\cite{Antipin:2019vdg}. Summing over their results, we obtain
\bsub
\label{eq:hyperdata}
\beq
\expec{\ga}^\text{hypercubic}_r = \frac{2r(r-1)(N-1)}{3N(N+1)}
\qaq
\expec{\ga^2}^\text{hypercubic}_r = \frac{r(r-1)(N-1)}{9N^2(N+1)} \; \mca{U}(N,r)
\eeq
with
\begin{align}
  \mca{U}(N,2) &= (N+2)^2(N+3)\\
  \mca{U}(N,3) &= (N+3)(N^2 + 22N-8)\\
  \mca{U}(N,4) &= N^3+51N^2 + 126N - 88\\
  \mca{U}(N,5) &= N^3 + 85N^2 + 220N - 180.
\end{align}
\esub
For reference, the hypercubic fixed point has
\beq
\ba_0(\la) = \frac{2}{3}(N-1)
\qaq
|\la|^2 = \frac{(N-1)(N+2)}{9N}.
\eeq

\section{Proof of equation~\reef{eq:sumrules}}
\label{sec:proofsumrules}

Let us first turn to the proof of~\reef{eq:sumrules}. We will first choose a basis $\{\mbf{e}_\a\}$ of symmetric tensors of rank $r$, normalized such that
\beq
\label{eq:basisNorm}
\mbf{e}^{\a}_{i_1 \dotsm i_r} \mbf{e}^\b_{i_1 \dotsm i_r} = \dd_{\a\b}
\quad
\Leftrightarrow
\quad
\sum_{\a=1}^{d_{N,r}} \mbf{e}^{\a}_{i_1 \dotsm i_r} \mbf{e}^\a_{j_1 \dotsm j_r} = \frac{1}{r!}(\dd_{i_1 j_1} \dotsm \dd_{i_r j_r} + \text{permutations}).
\eeq
Working in this basis, the anomalous dimensions $\{\gamma_\a\}$ are eigenvalues of the $d_{N,r} \times d_{N,r}$  matrix
\beq
\frac{r(r-1)}{2}\, M_{\a\b}(\la),
\qquad
M_{\a\b}(\la) \ldef \la_{ijkl} \, \mbf{e}^\a_{ijm_1 \dotsm m_{r-2}} \mbf{e}^\b_{kl m_1 \dotsm m_{r-2}}.
\eeq
In particular, we can express sums over eigenvalues as
\beq
\label{eq:sum11}
\expec{\ga^k}_r^\la = \frac{1}{d_{N,r}} \left(\frac{r(r-1)}{2}\right)^k \, \text{tr}\, M(\la)^k.
\eeq
Such traces are easiest to evaluate in an index-free formalism. For the $k=1$ case, let us introduce auxiliary vectors $u^i$ resp.\@ $v^j$, such that 
\beq
\text{tr}\, M(\la) = \frac{\la_{ijkl}}{(r!)^2} \frac{\pd^4}{\pd u^i \pd u^j \pd v^k \pd v^l} \, (\mca{D}_{u,v})^{r-2}\, (u \cdot v)^r,
\quad
\mca{D}_{u,v} \ldef \frac{\pd^2}{\pd u^i \pd v_i}.
\eeq
This is nothing but a rewriting of the second equation of~\reef{eq:basisNorm}. Applying the operator $\mca{D}_{u,v}$ $r-2$ times, we find
\beq
(\mca{D}_{u,v})^{r-2} \, (u \cdot v)^r = \frac{(r!)^2 d_{N,r}}{2N(N+1)}\, (u\cdot v)^2
\eeq
and performing the final four derivatives, we obtain
\beq
  \text{tr}\, M(\la) = \frac{d_{N,r}}{2N(N+1)}\,  \la_{ijkl} \, \frac{\pd^4}{\pd u^i \pd u^j \pd v^k \pd v^l}\,  (u\cdot v)^2  = \frac{2 d_{N,r}}{N(N+1)}\, \la_{iijj}.
\eeq
Inserting this into~\reef{eq:sum11}, the first equation of~\reef{eq:sumrules} is recovered.
For the second sum rule, we need to evaluate a slightly more complicated expression:
\begin{multline}
  \text{tr}\, M^2(\la) = \frac{\la_{ijkl}\la_{abcd}}{(r!)^4} \frac{\pd^4}{\pd u^i \pd u^j \pd w^k \pd w^l} \frac{\pd^4}{\pd v^a \pd v^b \pd t^c \pd t^d}\\
  \times (\mca{D}_{u,w})^{r-2} (\mca{D}_{v,t})^{r-2} \left[(u\cdot v)^r (w \cdot t)^r\right].
\end{multline}
Working out the second line, we get
\bsub
\beq
 \text{tr}\, M^2(\la) = \frac{d_{N,r}}{4r(r-1)(N)_4} \; \la_{ijkl}\la_{abcd} \, \frac{\pd^4}{\pd u^i \pd u^j \pd w^k \pd w^l} \frac{\pd^4}{\pd v^a \pd v^b \pd t^c \pd t^d} \, \mca{P}(u,v,w,t)
\eeq
where
\begin{multline}
\mca{P}(u,v,w,t) = (N+r)(N+r-1) (u\cdot v)^2 (w\cdot t)^2 + 4(r-2)(N+r) (u \cdot w)(u\cdot v)(w\cdot t)(v \cdot t) \\+ (r-2)(r-3) (u \cdot w)^2 (v \cdot t)^2.
\end{multline}
\esub
Computing the remaining derivatives, we obtain
\begin{align}
  \frac{1}{d_{N,r}}  \, \text{tr}\, M^2(\la) &= \frac{4}{r(r-1)(N)_4} \big[ (N+r)(N+r+1) \la_{ijkl}^2 \nn \\
    &\hspace{35mm} +4(r-2)(N+r) \la_{ijkk}^2 + (r-2)(r-3) \la_{iijj}^2 \big]\!.
\end{align}
Plugging this into~\reef{eq:sum11} with $k=2$, and using Eq.~\reef{eq:quad}, we finally obtain the second equation of~\reef{eq:sumrules}.

\section{Bound on $\expec{\ga}_r$ and $\expec{\ga^2}_r$ for general theories}
\label{app:specbnd}

In this section, we derive the bounds appearing in Eq.~\reef{eq:thmbounds}. Let's first obtain~\reef{eq:fact1} involving $\expec{\ga}_r^\la$. According to~\reef{eq:sumrules}, this quantity is proportional (by a positive constant) to $\ba_0(\la)$. But the latter quantity is bounded from above by its value at the $O(N)$ fixed point, as derived in Eq.~\reef{eq:a0bd}. Therefore~\reef{eq:fact1} follows immediately. 

Next, we will consider the bound~\reef{eq:fact2}, using Eq.~\reef{eq:supformula}. We will consider the cases $r=2$ and $r \geq 3$ separately. For $r=2$, the function $\mca{S}_{N,r} = \mca{S}_{N,2}$ reads
\beq
\mca{S}_{N,2}(x,y) = (N+2)(N+3)y = (N+2)(N+3)|\la|^2
\eeq
which only depends on $|\la|^2$ and not on $\ba_0(\la)$. So for this case, $\expec{\ga^2}_2$ can easily be bounded by appealing to~\reef{eq:RSref}.

For $r \geq 3$, the function $\mca{S}_{N,r}(x,y)$ depends both on $x=\ba_0(\la)$ and $y=|\la|^2$. Schematically, for fixed $N$, the domain $\mca{D}_N$ is of the form
\beq
\varrho_N^{-} < x \leq \varrho_N^+,
\quad
f_{-}(x) \leq y \leq f_{+}(x)
\eeq
where $f_\pm(x)$ and $\varrho_N^\pm$ depend only on $N$. The functions $f_\pm(x)$ are equal at $x = \varrho_N^\pm$, and we also know that $\varrho_N^+$ coincides with the $O(N)$ fixed point. The function $\mca{S}_{N,r}$ obeys
\beq
\frac{\pd}{\pd y} \, \mca{S}_{N,r}(x,y) = (N+r)(N - N_\star(r)),
\quad
N_\star(r) = 7r-17.
\eeq
For $N = N_\star(r)$, the function $\mca{S}_{N,r}$ does not depend on $y$; for $N > N_\star(r)$ the sup of $\mca{S}_{N,r}$ will be on the upper boundary curve $f_{+}$, whereas for $N < N_\star(r)$ it will be on the lower curve $f_{-}$.  

Let us first consider the case $N > N_\star(r)$. On the upper boundary curve, the function $\mca{S}_{N,r}$ is of the form
\beq
\mca{S}_{N,r}(x,f_{+}(x)) = \alpha_{N,r} \; x(\beta_{N,r} -x)
\eeq
for two constants $\alpha_{N,r}$ and $\beta_{N,r}$:
\beq
  \alpha_{N,r} = \frac{N^2-2 N r^2+4 N r+5 N-7 r^2+17 r}{2 N},
  \quad
  \beta_{N,r} = \frac{(N+r) (N+r+1)}{2\a_{N,r}}.
  \eeq
  If $\a_{N,r} > 0$ the maximum will be attained either at $x = \varrho_N^{-},\,\varrho_N^+$ or at $x = \beta_{N,r}/2$. The maximum will be attained at $\varrho_N^+$ iff $\beta_{N,r}/2 \geq \varrho_N^+$. For $r=3$, this requires that
  \beq
  r=3:
  \quad
  5 \leq N \leq 8 + 4 \sqrt{7} \approx 18.6.
  \eeq
  For $r \geq 4$, this condition is schematically of the form
  \beq
  r \geq 4:
  \quad
  \zeta'(r) \leq N \leq \zeta(r)
  \eeq
  where $\zeta(r), \zeta'(r)$ are two algebraic functions of $r$. The function $\zeta(r)$ will play a role later, and an explicit formula is included in the source code of this \TeX{} file.

 If $\alpha_{N,r} < 0$ then this function is a negative parabola, and it can only have a local \emph{minimum}. Therefore the sup of $\mca{S}_{N,r}$ will be attained at either $x = \varrho_N^{-}$ or $x = \varrho_N^+$. It turns out that the sup is always attained at $x = \varrho_N^+$. Hence $\mca{S}_{N,r}$ is also saturated by the $O(N)$ fixed point for
  \beq
  r \geq 4,
  \quad
  N_\star(r) < N < \zeta'(r).
  \eeq

  Next we consider the case $N = N_\star(r)$, where $\mca{S}_{N_\star(r),r}(x,y) \rdef \msc{S}_r(x) $ does not depend on $y$, and in fact
  \beq
  \frac{d}{dx} \msc{S}_r(x) = 4(r-2) \left[N_\star(r) + r + \th(r-3) x\right] > 0
  \eeq
  for all $r \geq 3$ and all allowed values of $x$. Therefore the maximum will be attained at $x = \varrho_N^+$.

  Finally we can consider the case $N < N_\star(r)$. In this case, we have to look for a supremum of the function $\mca{S}_{N,r}(x,f_{-}(x))$. The computation goes along the same lines as before. We find that for all $N < N_\star(r)$, the maximum is reached at $x = \varrho_N^+$.

  Summarizing, $\sup_{\mca{D}_N} \mca{S}_{N,r}$ is attained at $\varrho_N^+$, that is to say at the $O(N)$ fixed point, when
  \beq
  2 \leq N \leq N_\text{max}(r) \ldef \begin{cases}
    18 &\text{for} \quad r=3 \\
    \zeta(r) &r \geq 4
  \end{cases}\;.
  \eeq
Although $\zeta(r)$ is complicated, what is important is that the function $N_\text{max}(r)$ can be bounded as follows:
\beq
r \geq 3:
\quad
N_\text{max}(r) > 4\left(r-\frac{8}{9}\right)^2.
\eeq
This leads to the restriction on $N$ that appears in~\reef{eq:lastleg}.

\section{Bound for $\norm{g}$ at a complex fixed point}
\label{sec:complexvee}

The beta function~\reef{eq:complexbeta} can be written as
\beq
\label{eq:betarecast}
\beta(g)_{ijkl} \phib^i \phib^j \phi^k \phi^l = \left(- g_{ijkl} + (g \flat g)_{ijkl} + 4 (g\sharp g)_{ijkl}\right) \phib^i \phib^j \phi^k \phi^l
\eeq
where
\beq
\label{eq:music}
(g \flat h)_{ijkl} \ldef g_{ijmn} h_{mnkl}
\qaq
(g \sharp h)_{ijkl} \ldef g_{imkn} h_{jnlm}\,.
\eeq
In Eq.~\reef{eq:betarecast}, the effect of the contraction with $\phib^i \phib^j \phi^k \phi^l$ is merely to enforce the symmetrization according to~\reef{eq:compsymm}. Therefore, at a fixed point we have
\beq
\label{eq:initb}
\norm{g} \leq \norm{ g \flat g + 4 g \sharp g} \leq \norm{g \flat g} + 4 \norm{g \sharp g}.
\eeq
Now, for fixed $(i,j,k,l)$, we have
\beq
i,j,k,l\quad
\text{fixed}:
\quad
|(g \flat h)_{ijkl}|^2 =  \left| \sum_{mn} g_{ijmn}   h_{klmn}^* \right|^2 \leq \sum_{mn} |g_{ijmn}|^2 \sum_{pq}  |h_{klpq}|^2
\eeq
as follows by applying the inequality
\beq
\label{eq:CSCS}
|\text{tr}(A B^\dagger)|^2 \leq \text{tr}(A A^\dagger) \text{tr}(B B^\dagger)
\eeq
with $A_{mn} = g_{ijmn}$ and $B_{mn} = h_{klmn}$.  Consequently, $|g \flat h|^2 \leq \norm{g}^2 \norm{h}^2$. Likewise,
\beq
i,j,k,l\quad
\text{fixed}:
\quad
|(g \sharp h)_{ijkl}|^2 =  \left| \sum_{mn} g_{imkn} h_{lmjn}^* \right|^2 \leq \sum_{mn} |g_{imkn}|^2 \sum_{pq}  |h_{lpjq}|^2
\eeq
by applying~\reef{eq:CSCS} with $A_{mn} = g_{imkn}$ and $B_{mn} = h_{lmjn}$. Again, it follows that $\norm{g \sharp h}^2 \leq \norm{g}^2 \norm{h}^2$. In particular, for $g = h$ we conclude from~\reef{eq:initb} that
\beq
\norm{g} \leq 5 \norm{g}^2
\quad
\Rightarrow
\quad
g_{ijkl} = 0
\qor
\norm{g} \geq \tfrac{1}{5} \label{outputBK}
\eeq
as announced.